\documentclass[usenatbib,fleqn]{mnras}
\usepackage{amsmath}
\usepackage{lscape}
\usepackage[pdftex]{graphicx}     
\usepackage{aas_macros}
\usepackage[T1]{fontenc}
\usepackage{ae,aecompl}
\usepackage{amssymb}
\usepackage{soul}

\graphicspath{{.}}

\title[Post fall-back NS evolution and PSR activation]{Post fall-back evolution of multipolar magnetic fields and radio pulsar activation}
\author[A.P. Igoshev, J.G. Elfritz and S.B. Popov]{A.P. Igoshev,$^{1}$\thanks{E-mail: a.igoshev@science.ru.nl} J.G. Elfritz$^2$ and S.B. Popov$^3$\\
$^{1}$ Department of Astrophysics/IMAPP Radboud University Nijmegen P.O. Box 9010, 6500GL Nijmegen, The Netherlands\\
$^{2}$ Anton Pannekoek Institute, University of Amsterdam, Postbus 94249, 1090GE Amsterdam, The Netherlands\\
$^{3}$ Lomonosov Moscow State University, Sternberg Astronomical Institute, Universitetski pr. 13, 119991 Moscow, Russia }
\begin{document}

\date{Accepted 1988 December 15. Received 1988 December 14; in original form 1988 October 11}


\maketitle

\label{firstpage}

\begin{abstract}
 It has long been unclear if the small-scale magnetic structures on the neutron star (NS) surface could survive the fall-back episode. The study of the Hall cascade \citep*{cumming2004, wareing2009} hinted that energy in small scales structures should dissipate on short timescales. Our new 2D magneto-thermal simulations suggest the opposite. For the first $\sim$10~kyrs after the fall-back episode with accreted mass $10^{-3} M_\odot$, the observed NS magnetic field appears dipolar, which is insensitive to the initial magnetic topology.  In framework of the \cite{ruderman1975} vacuum gap model during this interval, non-thermal radiation is strongly suppressed.
After this time the initial (i.e. multipolar) structure begins to re-emerge through the NS crust. We distinguish three evolutionary epochs for the re-emergence process: the growth of internal toroidal field, the advection of buried poloidal field, and slow Ohmic diffusion. The efficiency of the first two stages can be enhanced when small-scale magnetic structure is present. The efficient re-emergence of high order harmonics 
might significantly affect the curvature of the magnetospheric field lines in the emission zone. So, only after few $10^4$~yrs would the NS starts shining as a pulsar again, which is in correspondence with radio silence of
central compact objects (CCOs). In addition, these results can explain the absence of good candidates for
thermally emitting NSs with freshly re-emerged field among radio
pulsars (\citealt*{bogdanov2014}), as NSs have time to cool down, and supernova remnants can
already dissipate. 
\end{abstract}

\begin{keywords}
stars: evolution - stars: neutron - stars: magnetic field - pulsars: general
\end{keywords}

\section{Introduction}\label{s:introduction}

Young neutron stars (NSs) are sources with wide ranges of characteristic observables, inferred fundamental parameters, and different energy supplies. The timing, rotational, cooling, and magnetic properties all provide hints to the coupled evolution (see \citealt{Harding2013} and references therein). The main classes of NSs include standard radio pulsars (PSRs), the soft gamma repeaters (SGRs) and anomalous X-ray pulsars (AXPs), the nearby cooling NSs called the ``Magnificent Seven'' (M7), the rotating radio transients (RRATs), and the central compact objects (CCOs) in supernova remnants (SNRs). 
The evolutionary scenario called the ``grand unification for NSs'' (GUNS) \citep*{guns, ipt2014}, attempts to explain this variety of sources.
Population synthesis studies have meanwhile yielded fruitful results. \cite{popov} described PSRs, magnetars, and M7 in a unified picture, and these results were later extended and improved by \cite{vigano2013} and \cite{gullon2014, gullon2015} with models covering larger ranges of parameter space, although CCOs were not included in these studies. The inclusion of CCOs have been discussed in a qualitative manner by \cite*{geppert}, but unification within the NS zoo remains problematic. A detailed population synthesis study of all known sub-populations of young NSs has not yet been successful, primarily because the birth process and subsequent evolution of CCO magnetic fields and observed emission remains an open question.

CCOs are young objects with typical ages of the order $10^4$\,yrs \citep{deluca2008}. These sources are characterized by relatively high surface temperatures and low dipolar (poloidal) fields (inferred from the spin-down period and period derivative). If these observable properties remain roughly unchanged on longer time scales, then we expect to see a significant population of low-field NSs in high-mass X-ray binaries (HMXBs), which have typical ages of $\sim$10$^6$ up to a few $10^7$\,yrs. However as it was demonstrated in a detailed study by \cite{cp2012}   no such systems had been observed. On the other hand, isolated CCOs could remain relatively hot until they have ages on the order $10^5$\,yrs, but such sources are not observed among nearby cooling NSs \citep{turolla2009}, which implies that CCOs might ``disappear'' after $\sim$10$^5$\,yrs. The explanation for this disappearance was found in the scenario of field re-emergence after fall-back. Soon after a SN explosion, a significant amount of infalling material can blanket the NS surface due to the reverse shock \citep{chevalier}, which can bury the NS magnetic field if the amount of material in the fall-back episode exceeds $\sim$10$^{-4}$~-~$10^{-3} \,\mathrm{M_\odot}$ (which is a function of the magnetic field; see e.g. \citealt*{bern10}). Following this process, the magnetic field then diffuses slowly back to the NS surface on a time scale $\sim$10$^4-10^5$\,yrs \citep*{ho11, vigano2012,bern12}. This screening can also be effective for large surface magnetic fields, leading to the so-called ``hidden magnetar'' scenario \citep*{hidden}. Several examples of such objects have been proposed, such as the NS in SNR Kes 79 \citep{sl12} and the NS in RCW103 \citep*{rcw103}.

As the buried field diffuses through the crust towards the NS surface, one would expect to see an active magnetar or a PSR (\citealt{geppert}, \citealt{vigano2012}). This process is also known as magnetic field re-emergence. In this case, such PSRs would be observed to have a growing dipolar field component.
 \cite{bogdanov2014} and \cite{luo2015} conducted searches for evolved CCOs which would be observed as PSRs, but no such sources have been detected. It is critical to explain the null results of these searches for evolved CCOs among radio pulsars.
 
 In  an earlier study, \cite{vigano2012} focused mostly on re-emergence of the dipole component. It is an oversimplification to think that the growth of dipolar component alone
 is enough for activation of pulsar emission. On the contrary, the weak dipolar magnetic field does not generally prevent NS from emitting non-thermal radiation (millisecond pulsars).
Therefore it is essential to choose a condition which allows us to distinguish between active and dormant PSRs. The usual criterion for pulsar activity is the proximity of a source to the death line in the period -- period derivative diagram. 
Studies of the pulsar ensemble consider three different aspects of the death line: observational\,--\,absence of known pulsars in the right lower corner of the period -- period derivative diagram; theoretical\,--\,a drop in electric potential at the magnetic polar cap \citep{ruderman1975}; and populational\,--\,pile up of pulsars near the death line \citep{bhattacharya1992, gonthier2002}.  The general consensus is that all three of these aspects correspond to the same physical mechanism, but this conclusion is not final yet.

\begin{figure}
\includegraphics[width=84mm]{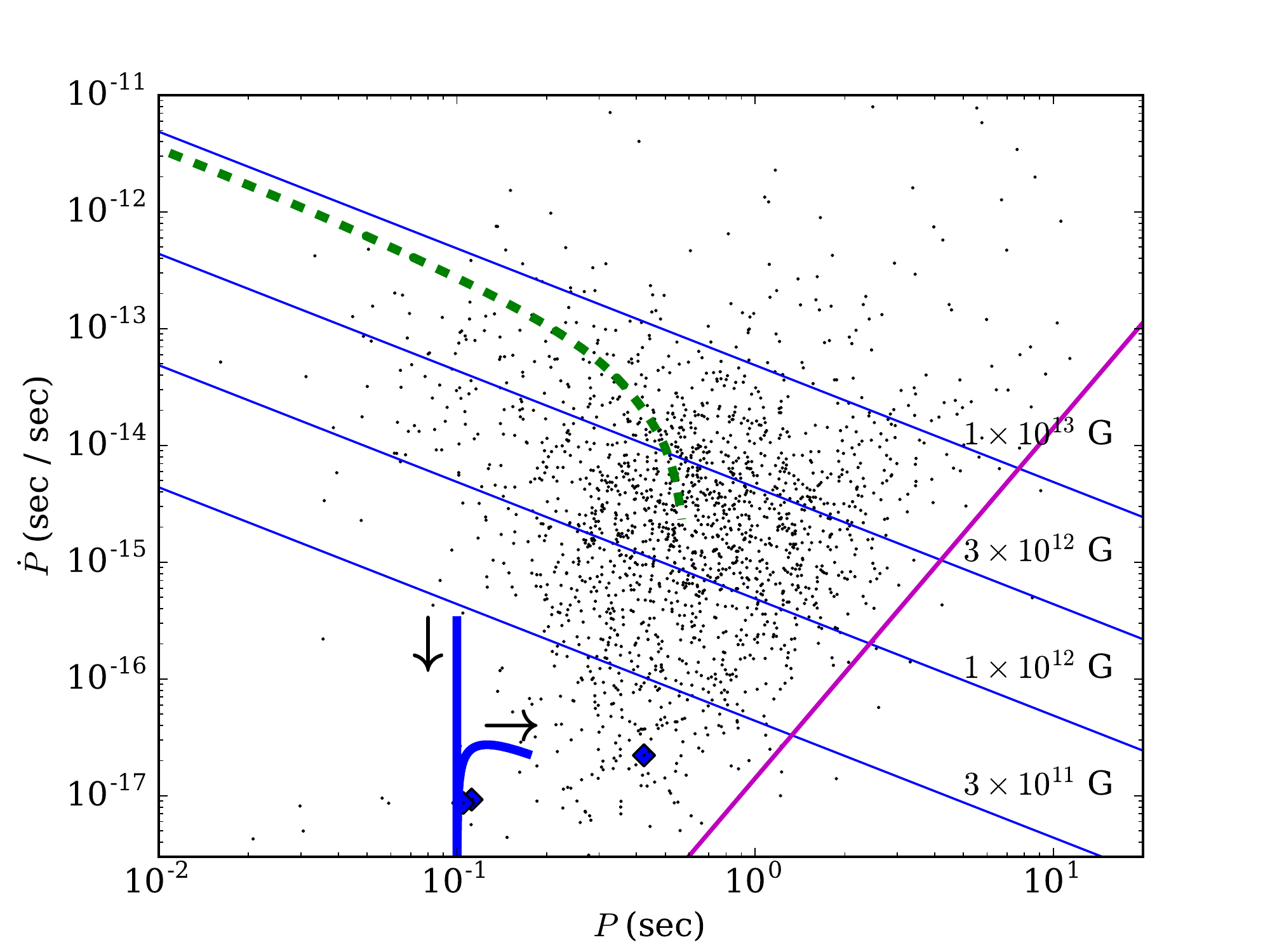}
\caption{$P$\,--\,$\dot P$ distribution for normal radio pulsars from the ATNF catalogue \citep{atnf}, and two evolutionary tracks for (a) a non-accreting NS with dipolar magnetic field and aligned quadrupole (dashed green line), and (b) a NS with a pure dipolar magnetic field which was buried with $\dot M = 10^{-3} \mathrm{M_\odot}$\,yr$^{-1}$ during a one year accretion phase (solid blue line). CCOs with measured period and period derivatives are shown as blue diamonds \citep*{halpern2010,gotthelf2013}. Black arrows indicate the direction of time, magenta solid line indicates the death line \citep{bhattacharya1992}.}
\label{f:p_dotp}
\end{figure}

From the ATNF catalogue\footnote{\texttt{http://www.atnf.csiro.au/research/pulsar/psrcat}} \citep{atnf}, the observed ensemble of normal radio pulsars is not strictly bounded to the right in the $P-\dot{P}$ diagram. A few pulsars lie below this line (see Fig.\,\ref{f:p_dotp}). One such source, J2144-3933, has a spin period 8.5\,s \citep*{young1999}. Although spin-down magnetic fields for CCOs measured by period $P$ and period derivative $\dot{P}$ are small, these objects are still well above the observed death line and should therefore shine as radio pulsars. The theoretical death line based on a pure dipolar magnetic field crosses the centre of the pulsar distribution \citep{medin_lai2007} and places some of the CCOs to the pulsar graveyard. To move the death line to its usual location, it is necessary to assume either a misalignment between the rotational and magnetic axes, or the presence of small-scale magnetic fields \citep{ruderman1975}. In pulsar population studies the pile up appears when neither luminosity nor magnetic field decay are assumed, and pulsars spend most of their lives in the region with characteristic ages $\sim$10$^8$\,yrs with nearly constant $P$ and small $\dot P$, thereafter observing the paucity of these sources. \cite{Szary2014} suggested to consider a limit on radio efficiency which helps to avoid pile-up as well. 

In the recent study by \cite*{szary2015}, the small scale magnetic field is modelled with additional dipoles which allows to vary the curvature radius for the open field lines in large range $\sim$10$^5$~--~$10^8$\,cm. Subject to strong fall-back, the magnetic pole may be shifted to a new orientation. Therefore, we consider not just an additional small dipole at the polar cap region as it was suggested by \cite{szary2015}, but rather harmonics of high order which can provide the necessary radius of curvature in the NS emission zone.

There is an important theoretical uncertainty concerning evolution
of  small-scale magnetic structures in NS
after a fall-back episode. The analysis of the Hall cascade properties
 \citep{cumming2004, wareing2009}
provided arguments that energy in small
scales structures should dissipate rapidly. Here we address this
question with numerical modelling.
In this paper we extend the first analysis performed by \cite{vigano2012} in two important aspects: we analyze and present evolution of high-order multipoles and study the non-thermal emission criterion to obtain an answer to the question if NSs with freshly re-emerged magnetic fields can be observed as radio pulsars or not. 

In \S\ref{s:method} we briefly describe the numerical model used to conduct our simulations of magnetic field burial and evolution. In \S\ref{s:results} we discuss simulation results employing a set of different initial conditions which vary in their prescribed multipolar structure. \S\ref{s:theory} contains an explanation of the numerical results in the context of the Hall cascade given our choice of initial conditions. Consequences for observable NS emission are presented in \S\ref{s:emission}. Limitations of both the emission model and the numerical code are discussed in \S\ref{s:discussion}, along with applicability to pulsar searches. We present our conclusions in \S\ref{s:conclusions}.

\section{Method}\label{s:method}

We perform our numerical experiments using two dimensional (2D) magneto-thermal simulations which self-consistently evolve the coupled magnetic field and the temperature throughout the NS interior. The numerical model is the popular code developed in \cite*{2012CoPhC.183.2042V}, which is based on the earlier model of \cite*{Pons2009}. The magnetic induction equation and the thermal evolution equation are coupled in non-trivial ways via the temperature-dependent electrical conductivity \citep*{Aguilera2008}, and it is necessary to evolve these equations as a system in order to correctly reproduce observed NS emission \citep{Pons2009,rea2012,vigano2013}. A brief overview of our numerical scheme is in order, although the interested reader is directed to these earlier studies for a more detailed treatment.

First, we prescribe an appropriate magnetic induction equation. We consider Hall-Ohmic evolution but also impose an advective electric field in our generalized Ohm's Law, such that the full induction equation is

\begin{equation}\label{eq:MTinduction}
\frac{\partial \vec B}{\partial t} = - \vec \nabla \times [ \eta \vec \nabla \times \vec B + f_h (\vec \nabla \times \vec B) \times \vec B + \vec{v}_{\mathrm{accr}} \times \vec B].
\end{equation}

\noindent Here $\eta=c^2/4\pi\sigma$ is the magnetic diffusivity, and $\sigma$ is the local, time-dependent electrical conductivity computed in the electron relaxation approximation. The time-independent Hall prefactor is $f_h = c/4\pi e n_e$ and has only radial dependence via the local electron density $n_e$. The velocity $\vec{v}_{\mathrm{accr}}$ allows us to manually bury the magnetic field during the accretion epoch, and takes the algebraic form

\begin{equation}\label{eq:vaccr}
\vec{v}_{\mathrm{accr}} = -\frac{\dot m}{4\pi r^2\rho(r)} \hat r,
\end{equation}

\noindent where $\rho(r)$ is the local mass density in the crust. The accretion epoch is $t_{\mathrm{accr}}=1$\,yr, beginning at $t=0$, during which period $10^{-3}$\,M$_{\odot}$ of material is accreted uniformly onto the NS surface; thus $\dot m=10^{-3}$\,M$_{\odot}$\,yr$^{-1}$. As already reported by \cite{TorresForne2016}, this accretion rate is not enough to bury $10^{12}$\,G fields on $\sim$1\,yr time scales\footnote{If the accretion rate drops as $\propto t^{-5/3}$, which was suggested by \cite{chevalier}, the instant accretion rate in the first few hours is much larger than $10^{-3}$ M$_\odot$/yr. This makes the submergence process possible while keeping the total amount of accreted matter low enough to neglect its influence on the crustal composition.}. However we are concerned with the first Megayear (Myr) of evolution after the field is buried, and not with modelling the transition from the proto-NS era onward. \cite{chevalier}, \cite{bern10} and \cite{bernal2016} have already carefully reported on the huge instant accretion rates associated with the hyper-critical accretion phase, where $\sim 100$\,M$_{\odot}$\,/yr can fall back onto the NS surface. Accurately capturing thermal evolution during this era requires understanding the compositional evolution, and how the accreted matter affects e.g. rapid neutrino cooling. However, linking the physical conditions during the supernova explosion to the subsequent hyper-critical phase, and thus to our initial conditions, is beyond the scope of this article. Our phenomenological model in Eq.\,(\ref{eq:vaccr}) is valid if we assume the hyper-critical accretion phase has already passed, the NS crust has formed, and thus the internal NS structure has been fixed. Thus we consider our Eq.\,(\ref{eq:vaccr}) sufficient to capture the essential secular signatures of the magnetic re-emergence process. We choose the total accreted mass to be $10^{-3}$\,M$_{\odot}$. Larger total accreted mass can bury the field completely (see e.g. fig.\,2 in \cite{hidden} for 0.01 M$_\odot$), while a smaller total accreted mass can lead to very fast reemergence, which is inconsistent with current CCO observations.

We employ a Skyrme type equation of state (EOS) with SLy nuclear interactions \citep{Douchin2001}. For densities below neutron drip point we also use the BPS relation \citep*{Baym1971}. We assume a nominal NS mass of 1.40\,M$_{\odot}$, and our EOS provides a star with radius $R_{NS} = 11.503$\,km and a core radius of $R_c = 10.799$\,km. We restrict our study to fields confined to the NS crust, and consider the NS core to be an ideal superconductor, thus we enforce as inner boundary conditions that tangential electric fields vanish at the crust-core interface ($E_{\theta}=E_{\phi}=0$); Ohmic dissipation timescales are $\sim$10$^{9}$\,yrs in the core, and dynamics in the NS core are effectively de-coupled from fundamental observables driven in NS crust \citep{Elfritz2016}, so it is reasonable to ignore the NS core in this study. At the $r=R_{\mathrm{NS}}$ outer boundary we decompose the radial component of the surface field to construct a multipole spectrum, consistent with a potential solution, and valid for both vacuum and force-free magnetospheres (\citealt*{Gralla_2016}).


\begin{figure*}
\begin{minipage}{0.48\linewidth}
\includegraphics[width=84mm]{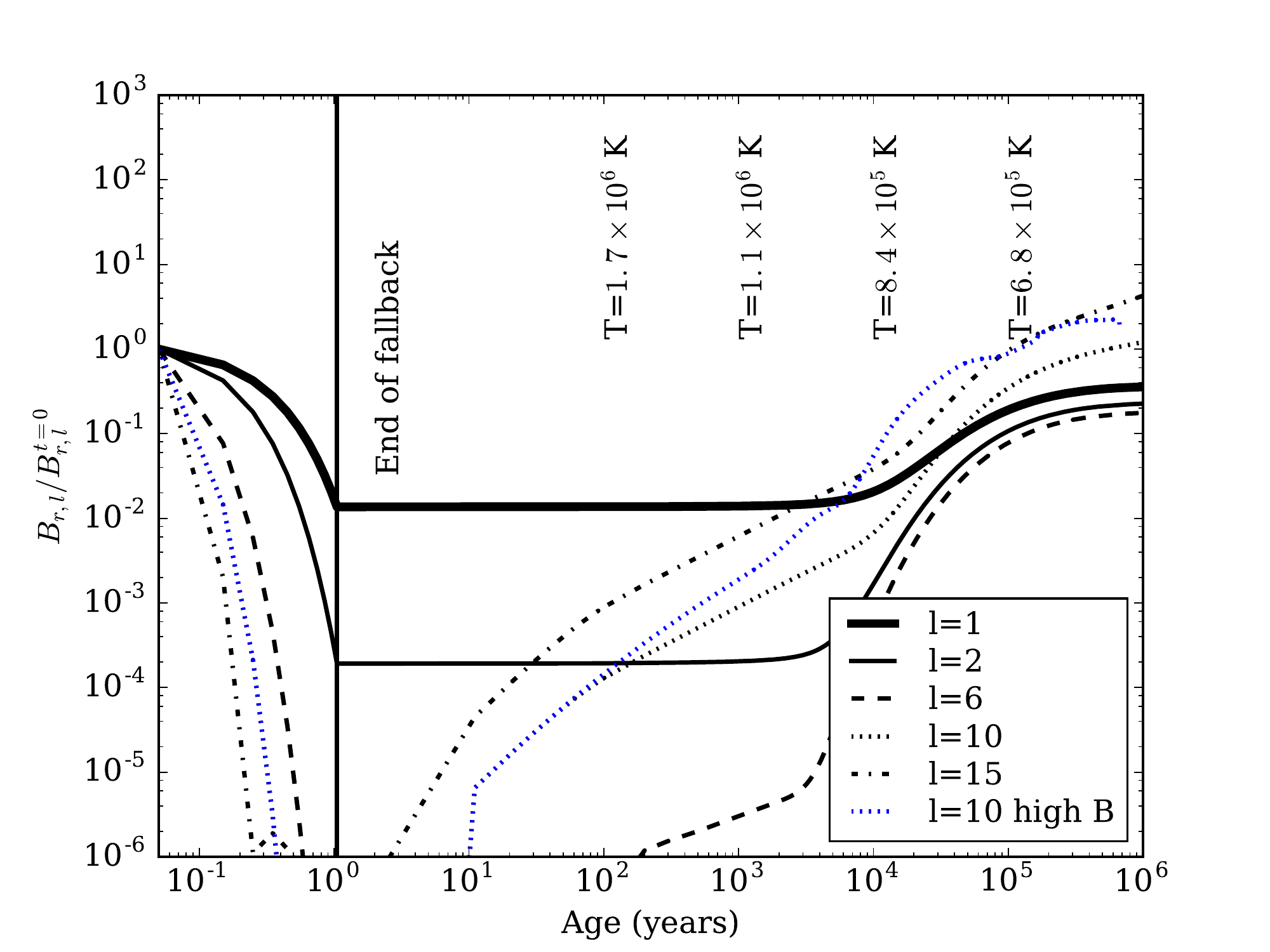}
\end{minipage}
\begin{minipage}{0.48\linewidth}
\includegraphics[width=84mm]{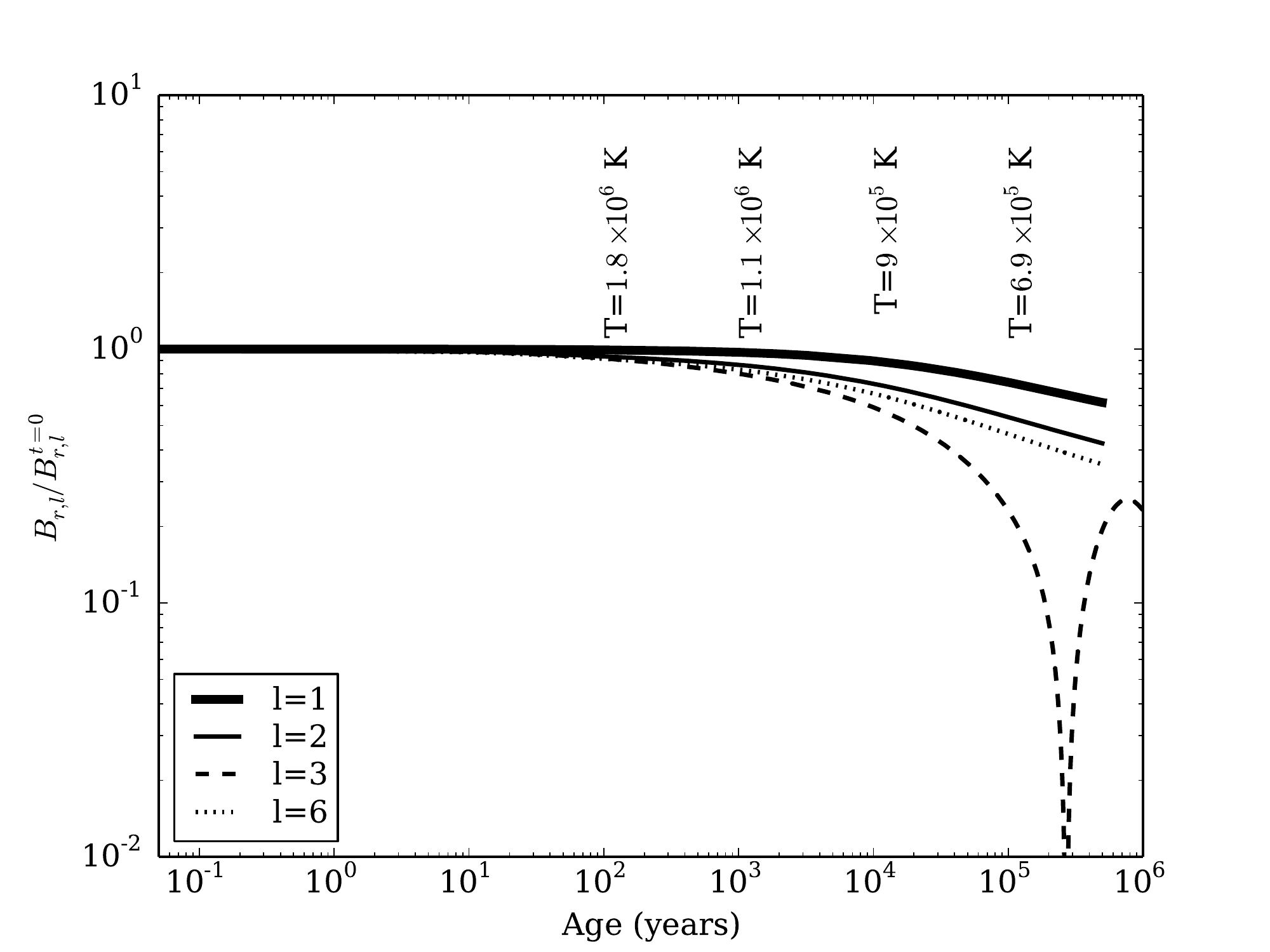}
\end{minipage}
\caption{Evolution of initially existing components of the surface poloidal magnetic field. {\it Right panel}: no fall-back. {\it Left panel}: fall-back with the total accreted mass $\delta M=10^{-3}\,\mathrm{M_\odot}$. In each panel several models with different initial field structure are shown. In every model the field initially consists of the dipole and one additional component. Evolution of the dipole component ($l=1$, thick solid line) is nearly identical for all models in each panel, and so just one curve for $l=1$ is shown in both panels. Other curves in each panel correspond to evolution of different components, which initially existed together with the dipole. Labels in legends in each panel mark the component which existed in the inital conditions, and which evolution is shown. Other components in all models, which were not present in the initial conditions and are generated during field evolution, are significantly weaker and are not shown here. The full map of surface multipole expansion is presented for several models in Fig.\,\ref{f:dyn_spectrum}. End of the fall-back period is marked in the left panel. In both panels we show several values of surface temperature at different moments. Temperature evolves in identical way in all models in each panel. The blue line in the right panel corresponds to a higher initial field strength of $3\,\times\,10^{13}\,$G.
}\label{f:l2_6_10}
\end{figure*}

The specific determination of our initial conditions is discussed separately in Appendix \ref{s:appendixICs}, but simply put, we prescribe superpositions of multipolar magnetic fields at $t=0$. We impose a background purely poloidal dipole, and superpose additional multipoles of order $l$. The initial toroidal field component is taken to be zero. The total magnetic field intensity at the north pole is normalized to $B_{\mathrm{pol}} = 1.5\times 10^{12}$\,G, with the dipolar component contributing $0.5\times 10^{12}$\,G. 

The thermal evolution is computed by the thermal transport equation for local temperature $T$,

\begin{equation}\label{eq:MTtemperature}
c_v \frac{\partial T}{\partial t} - \vec \nabla \cdot [\hat{\kappa} \cdot \vec{\nabla}T] = \sum_i Q_i
\end{equation}

\noindent where $c_v$ is the local specific heat capacity and $\hat{\kappa}$ is the thermal conductivity tensor. In the right side of Eq.\,(\ref{eq:MTtemperature}) we assume no additional sources of heat during the accretion phase for two reasons: first, the accretion process is very short (duration of 1\,yr in our simulations) and second, the accretion stage is at the very beginning of the NS evolution when the star is still extremely hot ($\sim$10$^{10}$\,K), and thus any accretion-induced heating due to infalling material is quickly quenched by neutrino emission. Therefore the only non-zero $Q_i$ that we must include are the Joule heating and usual neutrino cooling terms. Our code traces the thermal and magnetic field evolution up to magnetar field strengths, therefore all relevant energy sinks via neutrino cooling are included (see Table 4.3 of \citealt{vigano_thesis}). 

In order to focus on the field evolution during the burial and re-emergence phases, we reduce the inner crust impurity factor to 0.1, such that associated dissipation is suppressed. We use a numerical grid with $n_r=30$ radial cells through the NS crust, and adjust the number of cells in polar angle to suit the angular variation of the imposed $l$\,-pole (varies from 50 - 300 cells). 1\,Myr is chosen as the simulation time in each case. 

\section{Results: the re-emergence process}\label{s:results}

In this section we describe the results from our magneto-thermal simulations for a variety of initial conditions. We compare re-emergence of the imposed multipolar fields, shown alongside identical simulation results without the initial accretion phase (Fig.\,\ref{f:l2_6_10}). 

We start the discussion of the magneto-thermal evolution from a case with no fall-back which was already presented multiple times in the literature, to verify that our results are consistent. The low order harmonics ($l=1, 2$) decay on the Ohmic timescale, about 1 Myr, see right panel of Fig.~\ref{f:l2_6_10} \citep{cumming2004, pons2007}. Even though we assume no toroidal magnetic field in our initial conditions, this component is generated and typically saturates at values comparable to the initial poloidal field strength; that is, $B_{\phi}^{\mathrm{peak}}\approx\,0.5\,-\,2.0\,B_{\mathrm{pol}}^{t=0}$. The NS cools undisturbed and reaches a temperature around $6.9\times 10^5$~K at $t=10^5$ years, see temperature labels above the magnetic field curves at Fig.~\ref{f:l2_6_10}. This result is in agreement with \cite{Aguilera2008} for low-mass weakly magnetized NS. The high order harmonic ($l=6$) decays slightly faster than the dipole and quadruple.  

When the fall-back is introduced the evolution of low order multipoles ($l=1,2$) does not differ from the earlier published results \citep{vigano2012}: we found very similar reemergence time scales $\sim$10$^5$\,yrs, caused by diffusion of the poloidal magnetic field toward the surface. The surface magnetic field is reduced after the fall-back and the internal magnetic field is amplified because of compression by accreted matter \citep{bern12}. The surface temperature is not sensitive to the short fall-back episode. 
In simulations where high order harmonics ($l=6,10$ or $15$) are present with a background dipole, we find the shorter re-emergence time scale of $\sim$10$^4$ years. Moreover, the re-emerged field intensity of these higher order harmonics appears to be larger than what was imposed in the initial conditions. This result has important implication for the theory of pulsar emission and is discussed in details in \S\ref{s:emission}. During the first few hundred years a strong toroidal magnetic field is formed, up to $B_{\mathrm{tor}}\approx 7\times 10^{12}$\,G; see Figs.~(\ref{f:l10_crust},\ref{f:l15_crust}). This toroidal magnetic field has a large multipole number and survives during $10^6$ years, showing slow migration toward the crust-core interface. At the surface of the neutron star, additional harmonics are formed on the Hall time scale, see Fig.~\ref{f:dyn_spectrum}, their intensity though does not reach a significant value during the course of our simulations. The presence of high order multipole structure does not affect the bolometric temperature of the NS. 

If we choose initial conditions consisting of a dipole and $l=10$ harmonic with a higher magnetic field, $B=3\times 10^{13}$~G, we see very similar behaviour during the toroidal growth stage (compare blue and black dotted lines at left panel of Fig.~\ref{f:l2_6_10}), but the stronger Hall drift accelerates the re-emergence process somewhat. The surface multipole expansion Fig.\,\ref{f:dyn_spectrum} shows that more harmonics are generated, due to the Hall cascade, but their intensity is still small compared to the $l=1$ and $l=10$ field components.

We have also performed simulations which include non-zero toroidal field components at $t=0$, prior to the short fall-back, with the form

\begin{equation}\label{eq:toroidalform}
B_{\phi}=B_0\frac{R_0}{r}\left(\frac{2}{\Delta r_{\mathrm{cru}}}\right)^4\left(r-R_{\mathrm{NS}}\right)^2\left(r-R_c^2\right)\mathrm{sin}\,\theta,
\end{equation}

\noindent and found no significant difference in the surface multipole expansion in the first $10^6$\,yrs. Here $R_0$ and $\Delta r_{\mathrm{cru}}$ denote the radial center of the crust and the crustal thickness, respectively. $B_0$ normalizes the peak field strength. Our series of experiments have shown that with or without the inclusion of an initial toroidal component, the system will in general relax to the configurations shown in Figs.\,(\ref{f:l10_crust},\ref{f:l15_crust}) \citep{Gourgouliatos2014}.

\section{Theoretical explanation}\label{s:theory}

NS magnetic field evolution in the presence of high-order multipolar structure occurs in three distinct stages, the time scales of which are functions of both the Ohmic diffusion and Hall time scales; these are given by $\tau_{\mathrm{Ohm}}=4\pi\sigma L^2/c^2$ and $\tau_{\mathrm{Hall}}=4\pi e n_e L^2/cB$, respectively, where $L\approx 1$\,km is the characteristic length scale of field variation. The first evolutionary stage is the growth of toroidal field from the standard forward Hall cascade (starting from fall-back until $\sim$1\,$\tau_{\mathrm{Hall}}$), then advection of the poloidal field towards the surface of the neutron star caused by the poloidal-toroidal coupling in the Hall drift (typically up to $\sim$10$\tau_{\mathrm{Hall}}$), and finally the diffusion of poloidal field to the surface on the Ohmic time scale. The diffusive epoch depends primarily on the electrical conductivity $\sigma$ in the NS crust, and typically dominates the evolution beyond $\sim$10$^5$\,yrs for our chosen EOS. This final stage weakly depends on the angular structure of the field, but not the field intensity; this evolutionary stage is well-studied in the literature \citep{cumming2004, urpin1994}, and thus we treat only the first two epochs in this section.



\subsection{The linear toroidal field growth stage.}\label{s:torgrowth}

The early evolution is dominated by the induction of a toroidal magnetic field component, because we supply only poloidal components for the initial field. While an arbitrary toroidal component can easily be imposed (see \citealt{vigano2012}), the long-term NS evolution is not particularly sensitive to its inclusion. We specifically do not prescribe the $B_{\phi}$ components from Eq.\,(\ref{eq:finalbphi}) because this simply delays activation of the Hall cascade.

The non-dipolar components of the initial field, from Eqs.\,(\ref{eq:finalbr}, \ref{eq:finalbth}) are

\begin{equation}
B_{r,l} (r, \theta)= -\frac{\mu_l^2}{x^2} l(l+1) C_l \Gamma_l (x) P_l (\cos\theta)
\end{equation}

\begin{equation}
B_{\theta,l} (r,\theta) = -\frac{\mu_l^2}{x} C_l  \Gamma_l'(x) P_l'(\cos \theta)
\end{equation}

\noindent where $x = \mu_l r$ is a normalized radial coordinate. To satisfy boundary conditions we must take $\mu_1 = 2.29$, $\mu_2 = 0.628$, $\mu_3 = 0.313$ and all other $\mu_l \approx 0.273$\,km$^{-1}$ (see Appendix \ref{s:appendixICs}). 

In this early stage of the evolution, the toroidal field evolves according to the non-diffusive Hall induction equation


\begin{equation}
\partial_t B_\phi = -\vec{\nabla}\times\left[f_h (\vec{\nabla}\times \vec{B}_\mathrm{pol}) \times \vec{B}_\mathrm{pol}\right].
\end{equation}


\noindent For each multipole $l$ in the poloidal initial conditions, it can be shown that the corresponding azimuthal field grows according to


\begin{equation}
\partial_t B_\phi = - C_l^2  \mu_l^6 \frac{f_h}{x^6} [\psi(x) P_l P'_l  + \chi(x) P'_l P_l''],
\label{e:diff_bt}
\end{equation}

\noindent where we write the two radial functions $\chi(x)$ and $\psi(x)$ using the Bessel-Riccati differential equation and its derivative as

\begin{equation}
\chi(x) = -2x^3 \Gamma_l \Gamma_l'
\end{equation}

\begin{equation}
\psi (x) = 2xl (l+1) \Gamma_l [x\Gamma_l - l(l+1)\Gamma_l'].
\end{equation}

\begin{figure*}
\begin{minipage}{0.48\linewidth}
\includegraphics[width=84mm]{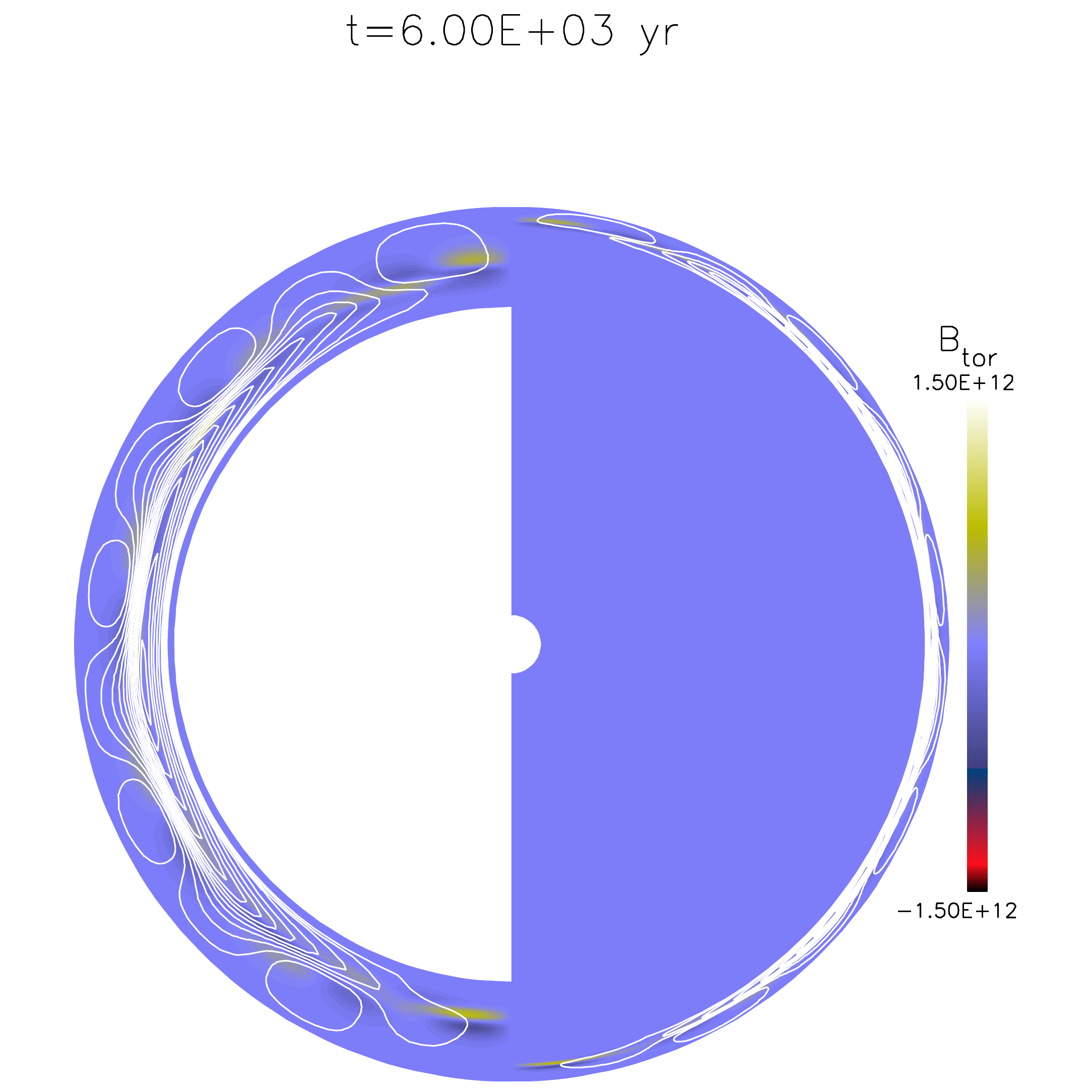}
\end{minipage}
\begin{minipage}{0.48\linewidth}
\includegraphics[width=84mm]{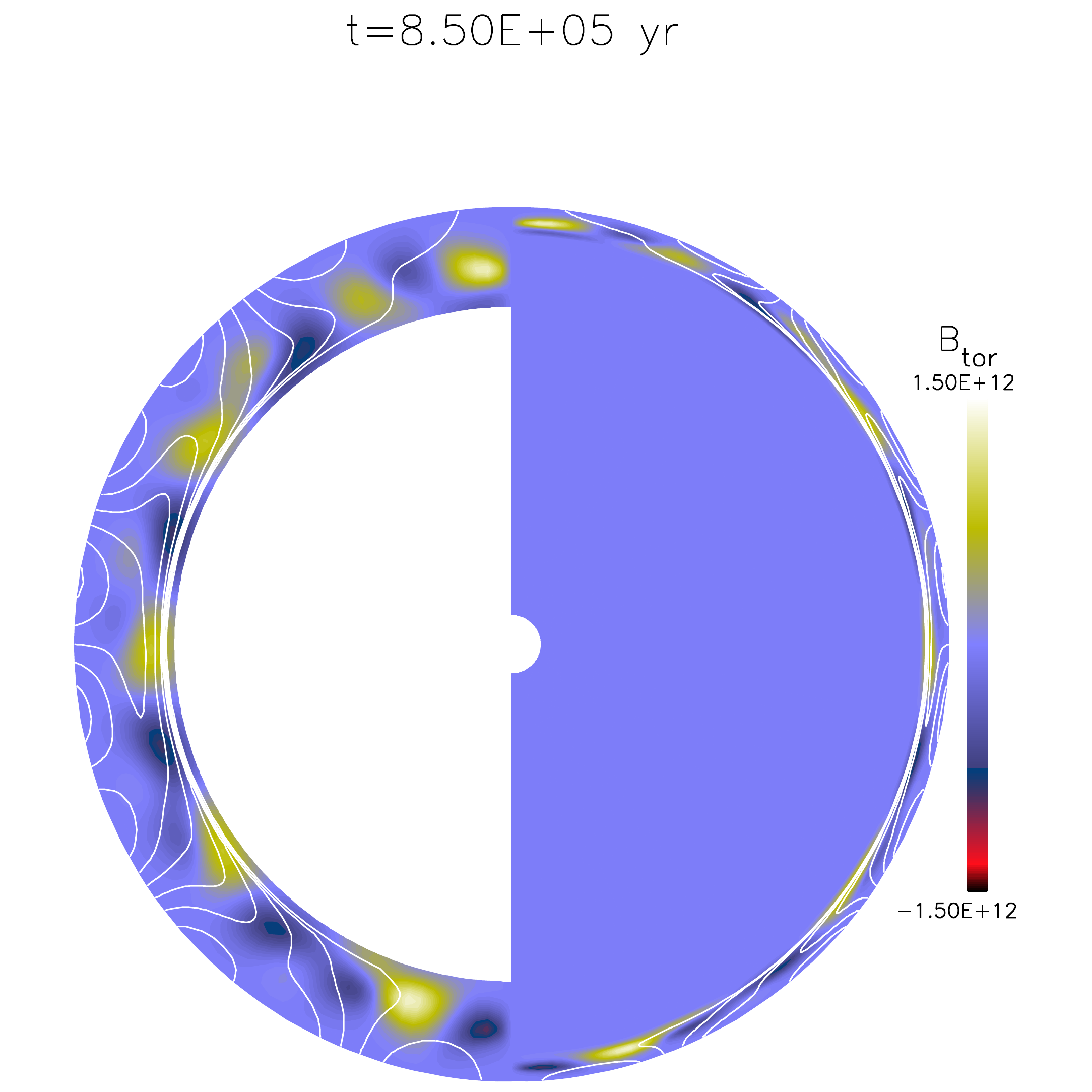}
\end{minipage}
\caption{
The crustal magnetic field configuration at two different times, for initial conditions consisting of dipole and $l=10$ harmonic. The crust is artificially enlarged in the left halves for clarity. White lines indicate field lines projected into the poloidal plane, and the background color indicates the toroidal field strength.
}
\label{f:l10_crust}
\end{figure*}

\begin{figure*}
\begin{minipage}{0.48\linewidth}
\includegraphics[width=84mm]{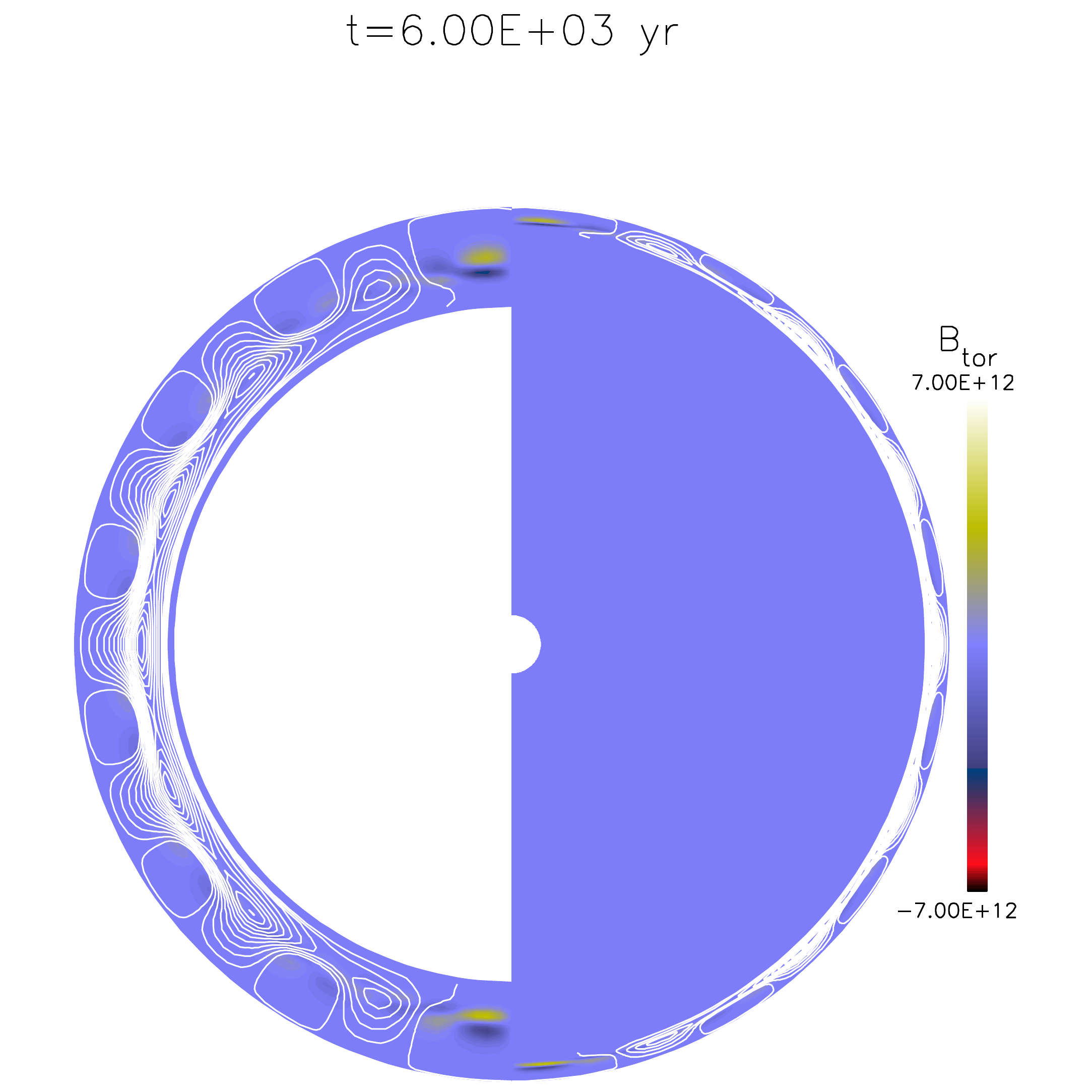}
\end{minipage}
\begin{minipage}{0.48\linewidth}
\includegraphics[width=84mm]{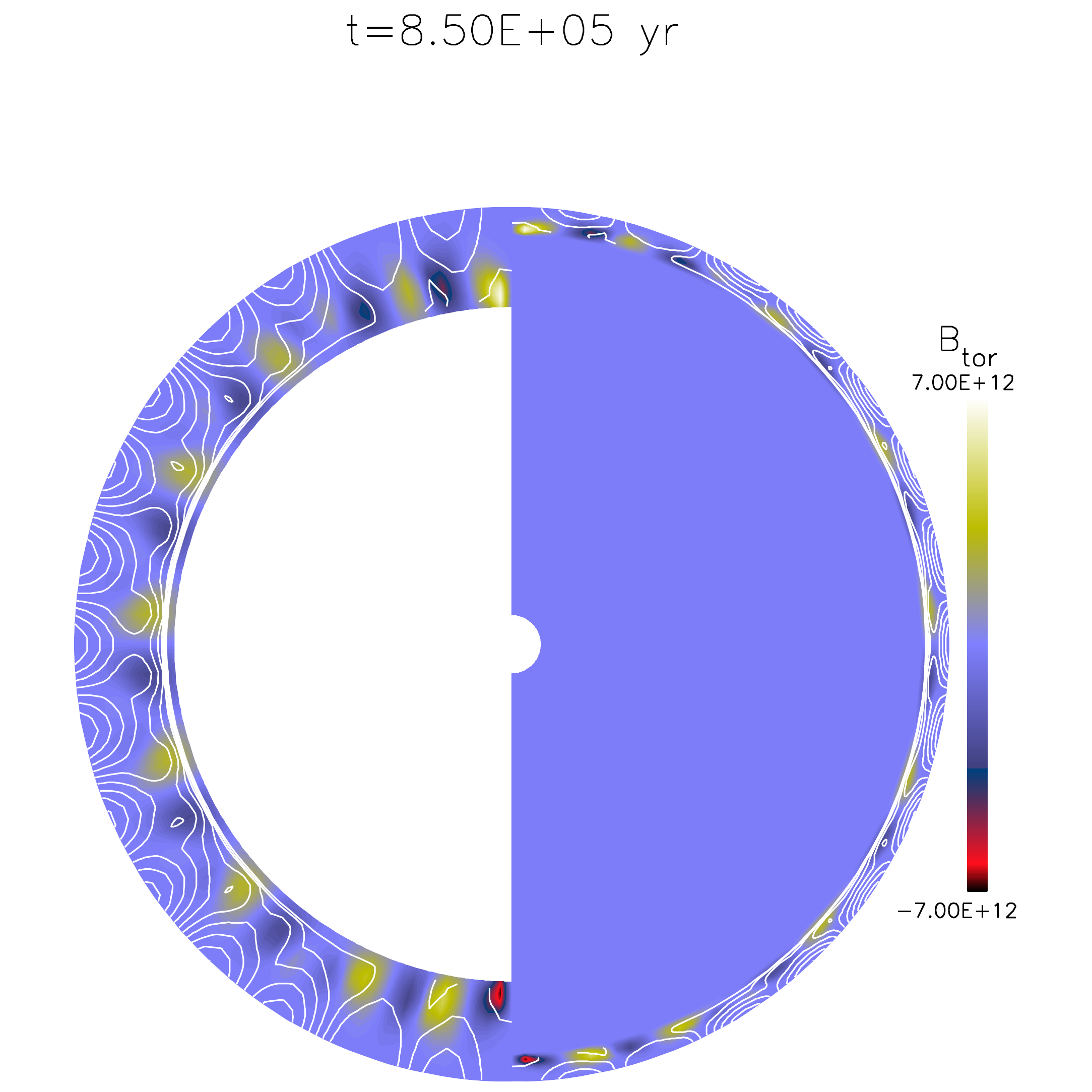}
\end{minipage}
\caption{
The crustal magnetic field configuration at two different times, for initial conditions consisting of dipole and $l=15$. The crust is artificially enlarged in the left halves for clarity. White lines indicate field lines projected into the poloidal plane, and the background color indicates the toroidal field strength.
}
\label{f:l15_crust}
\end{figure*}

From consideration of our boundary conditions, it is clear that $\Gamma_l(x)$ is a similar monotonic function for all multipoles, and is thus effectively independent of $l$. We then Taylor expand $\Gamma$ about the crust-core interface, keeping up to the linear term.
It follows that for typical values, $\Gamma_l\Gamma_l'\approx 2$ for high $l$. 
The second term in Eq.\,(\ref{e:diff_bt}) is the dominant contribution for the $l=1$ dipole case, whereas the first term dominates for all higher multipoles, due to the quartic dependence on harmonic number $l$. 
Thus for a given $l$, the fastest corresponding toroidal growth obeys

\begin{equation}\label{eq:bphisat}
\partial_t B_\phi \approx 4 C_l^2 f_h \frac{\mu_l^6}{x^5} l^2 (l+1)^2  P_l P'_l,
\end{equation}

\noindent illustrating that the toroidal field grows as the odd harmonics ($2l-1, 2l-3, 2l-5, ...$), due to the $P_l P_l'$ dependence.
It makes sense that the expected growth rate scales as $\sim$\,$1/l^2(l+1)^2$, since we impose angular structures of order $l$ as initial conditions, i.e., this is the scale size of our current sheets. 
The local timescale for large multipoles is therefore
\begin{equation}
\tau_{\mathrm{Hall}, l} \approx \frac{4\pi e n_e}{cB}\frac{L_0^2}{l^2 (l+1)^2}
\end{equation}
\noindent where $L_0$ is the length scale for the background, dipolar field. This analytic approximation is in good agreement with the growth shown in our numerical simulations. In the left panels of Figs.\,(\ref{f:l10_crust}, \ref{f:l15_crust}) we see the toroidal field during this linear growth phase for two large-$l$ cases. 
When the toroidal and poloidal field intensities become comparable, the toroidal growth saturates, and the next stage of evolution is controlled by the toroidal-poloidal coupling.


\subsection{The toroidal-poloidal coupling.}\label{s:torpol}

The toroidal magnetic field component is well developed after a few Hall time scales, which accelerates the evolution of the poloidal field from the coupling equation

\begin{equation}\label{eq:torpolcoupling}
\partial_t\vec{B}_{\mathrm{pol}}=-\vec{\nabla}\times\left[f_H\left(\vec{\nabla}\times\vec{B}_{\mathrm{tor}}\right)\times\vec{B}_{\mathrm{pol}}\right].
\end{equation}

\noindent As already shown by \cite{vigano_thesis}, the poloidal equations resemble standard advection equations which contain source terms quadratic in $\vec{B}$. We can write the relation in standard form, as

\begin{equation}\label{eq:torpoladvection}
\left(\partial_t +\vec{v}_e\cdot\vec{\nabla}\right)\vec{B}_{\mathrm{pol}} =\left(\vec{B}\cdot\vec{\nabla}\right)\vec{v}_{e,\mathrm{pol}} 
\end{equation}

\noindent where $\vec{v}_e=-f_H\vec{\nabla}\times\vec{B}$ is the electron velocity. Clearly the left-hand side of Eq.\,(\ref{eq:torpoladvection}) is an advective derivative on the poloidal field, while the right-hand side provides nonlinear feedback on the field depending on the configuration of the current system. The result is that the buried magnetic field advects towards the NS surface, and does so on faster time scales than for pure Ohmic diffusion. 



We can exploit the ordered multipolar structure in $\hat{\theta}$ to estimate the re-emergence time scale. There is no such periodicity in the radial direction, so we Fourier transform Eq.\,(\ref{eq:torpolcoupling}) in $\hat{\theta}$ in the limit $\nabla\rightarrow i\,\vec{k}_{\theta}$, where $\vec{k}$ is the usual wave vector. We also Fourier transform in time such that $\partial_t\rightarrow -i\,\omega$, and search for exponentially growing solutions. After some algebra, one may compute the re-emergence speed $v_{\mathrm{re-em}}$ via the group velocity, as

\begin{equation}
v_{\mathrm{re-em}}=\left|\frac{d\omega}{dk_{\theta}}\right|\approx v_0\cdot l\cdot B_{\phi}
\end{equation}

\noindent with $v_0=2 \,f_H/R_{\mathrm{NS}}$. This estimate shows that higher-order multipoles re-emerge at the NS surface earlier than e.g. a buried dipole component. There are important observational implications to this. The re-emergence time is then easily inferred based on the burial depth $\Delta\,r_{\mathrm{burial}}$: 

\begin{equation}
\Delta\,t_{\mathrm{re-em}}\approx\frac{\Delta\,r_{\mathrm{burial}}}{v_0\cdot l \cdot B_{\phi}}.
\end{equation}

In general the peak $B_{\phi}$ value at saturation\,--\,which is responsible for advecting the poloidal field to the surface\,--\,is comparable to the initial field strength $B_0=1.5\times 10^{12}$\,G. Using $R_{\mathrm{NS}}=11.5$\,km and $f_H\approx 2$\,km\,Myr$^{-1}$\,10$^{-12}$\,G$^{-1}$, with a 0.35\,km burial depth (half the crust thickness), it follows that the re-emergence time scale is roughly

\begin{equation}\label{eq:temerge}
\Delta\,t_{\mathrm{re-em}}\approx\frac{670}{l}\,\mathrm{kyrs}.
\end{equation}

\noindent This approximation agrees well with the re-emergence time scales found in Fig.\,\ref{f:l2_6_10}, within a factor of $\sim$2\,--\,5. 

\begin{figure*}
\begin{minipage}{0.48\linewidth}
\includegraphics[width=84mm]{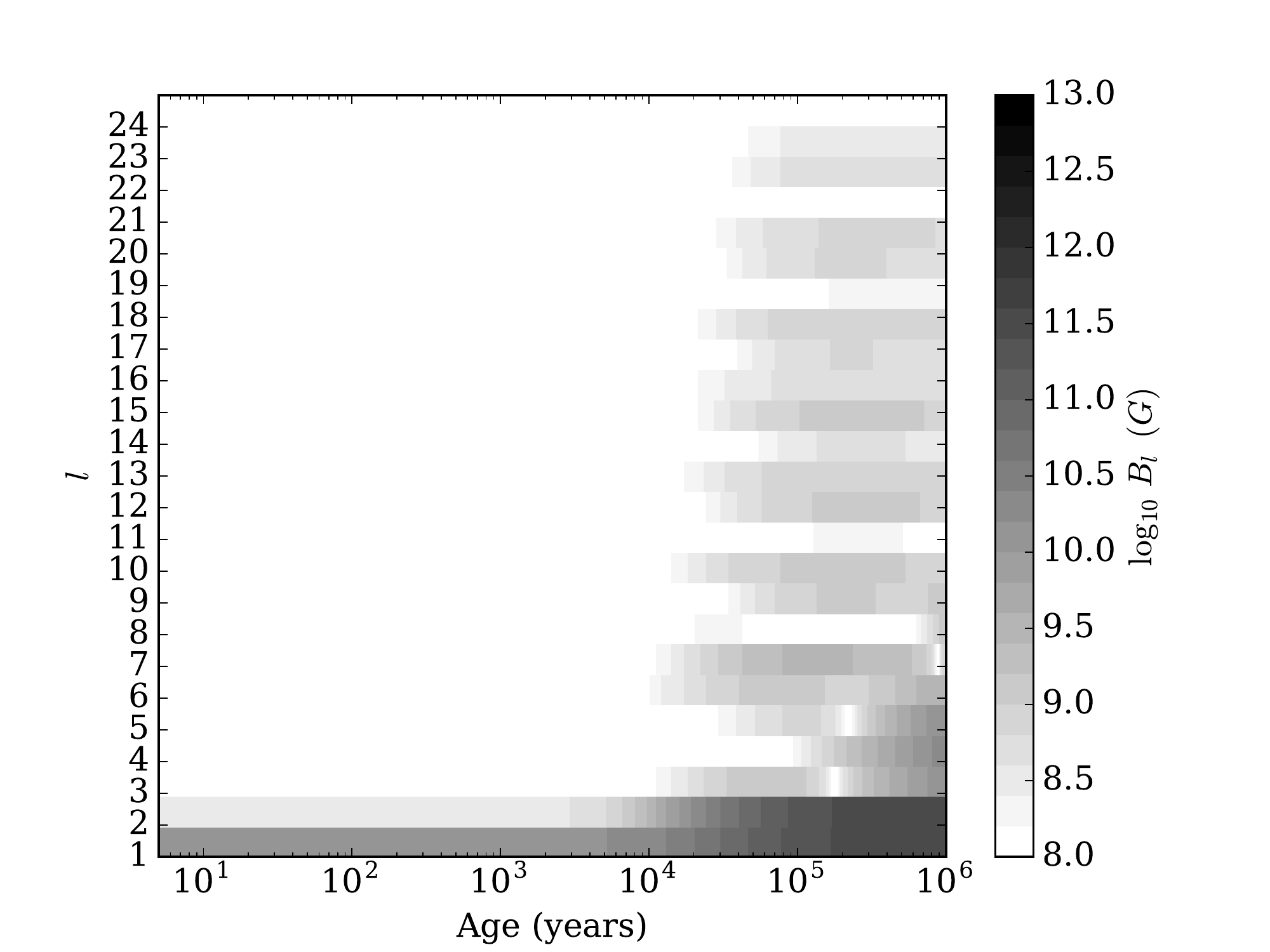}
\end{minipage}
\begin{minipage}{0.48\linewidth}
\includegraphics[width=84mm]{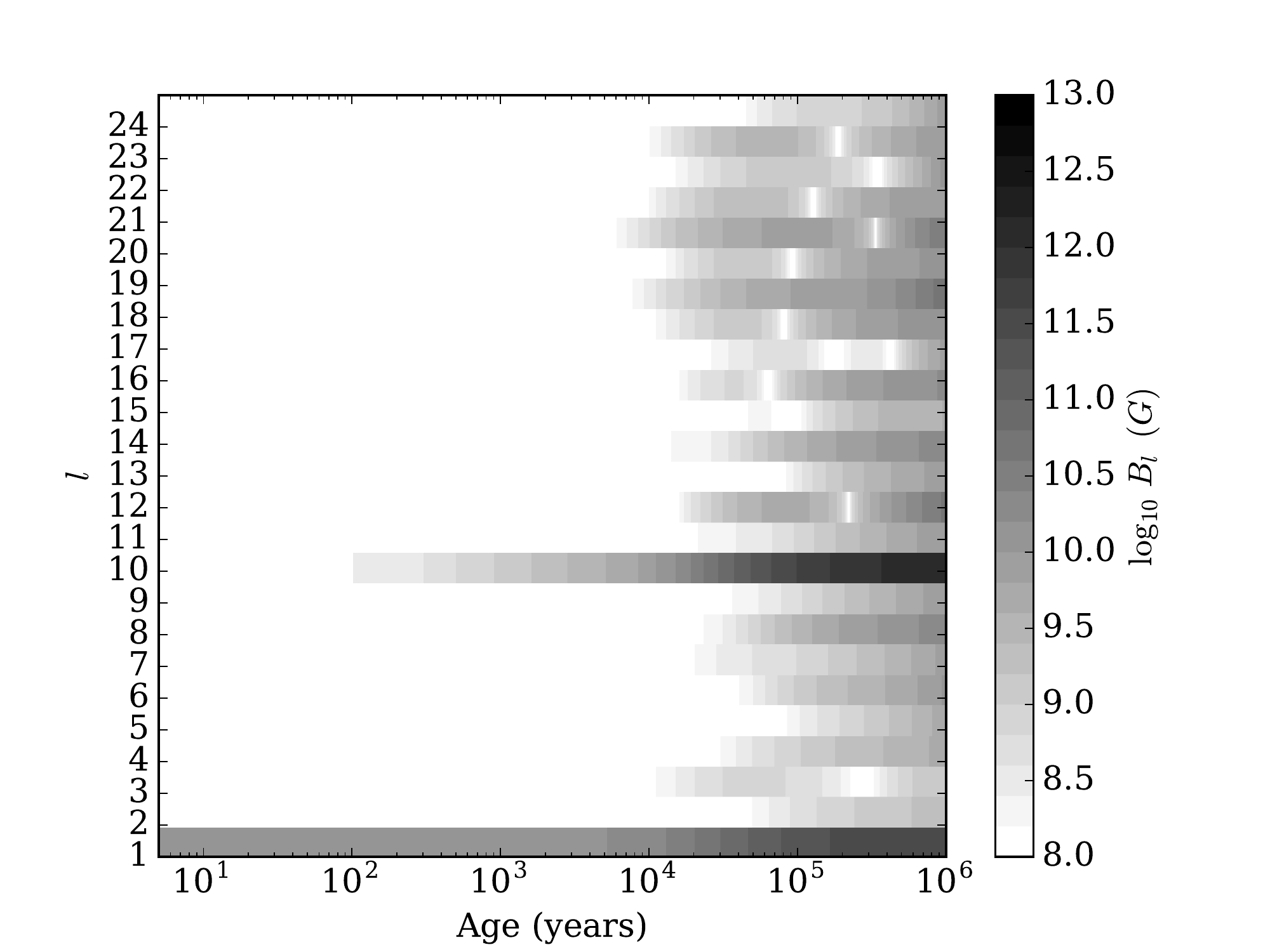}
\end{minipage}
\begin{minipage}{0.48\linewidth}
\includegraphics[width=84mm]{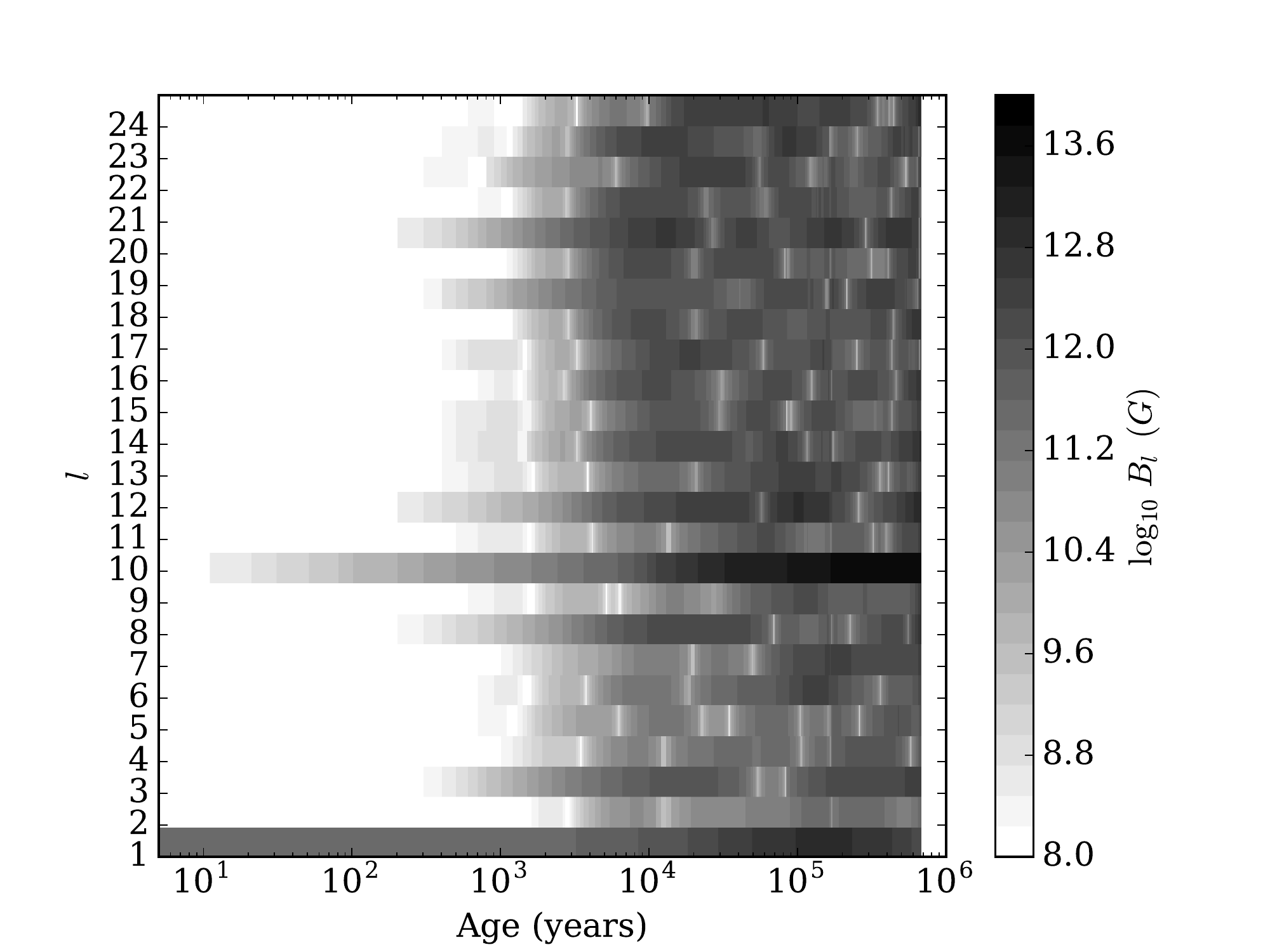}
\end{minipage}
\caption{
Surface poloidal magnetic field decomposition versus time (horizontal) and multipole number (vertical). The greyscale shows the strength of individual multipoles. {\it Top left}: the initial configuration consists of dipole and $l=2$ harmonic. {\it Top right}: the initial configuration consists of dipole and $l=10$ harmonic. {\it Bottom}: the initial configuration consists of dipole and $l=10$ harmonic with higher $t=0$ field strength.
}
\label{f:dyn_spectrum}
\end{figure*}

It is important to note that our choice of simulation inputs constrains the physics. The simulated re-emergence process is elastic, in the sense that the parity of re-emergent poloidal field is determined by the imposed initial conditions. In our experiments we have tested various combinations of odd-even and odd-odd harmonics (see \S\ref{s:results}), and for $t > \Delta\,t_{\mathrm{re-em}}$ the spectral decomposition at the NS surface strongly resembles the supplied initial conditions (Fig.\,\ref{f:dyn_spectrum}). The reason is that we are concerned with relatively weak field intensity at birth, $\sim$10$^{12}$\,G, and thus the nonlinear Hall drift is also somewhat weak. Therefore the Hall drift cannot efficiently accelerate Ohmic dissipation of the high-$l$ structure on sub-Myr time scales.

\section{Consequences for non-thermal emission}\label{s:emission}

Our objects have $B\sim$10$^{12}$~G and periods $P\sim$0.1-0.3\,sec, in correspondence with recent studies of initial periods (\citealt{popov2012}; \citealt{igoshev2013}). In this region of the $P$\,--\,$\dot P$ plane the efficiency of non-thermal emission is strongly controlled by the curvature radius of open field lines (see figs.\,8,9  in \citealt{medin_lai2007} for the position of the death line depending on the curvature radius). Curvature radiation is a dominant process which causes the cascade development and determines the emitted power, while the resonant Compton scattering does not play an appreciable role \citep{Timokhin_2015}.

Our results for field re-emergence, summarized in Fig.\,\ref{f:l2_6_10}, have immediate consequences for non-thermal emission of NSs. Sources subject to fall-back, shows primarily dipolar magnetic fields and ages up to $10^3$\,yrs, have large curvature radii in the open field line region ($\sim$100\,$R_\mathrm{NS}$), and thus produce negligible non-thermal radiation. After $\sim$10$^5$\,yrs, the large multipolar components dominate close to the surface and decrease the curvature radius, thus triggering non-thermal emission. This causes the neutron star to shine as a pulsar again.

We briefly describe the emission of pulsars (the interested reader is referred to \citealt{Timokhin_2010, Timokhin_2013, Timokhin_2015} for details) in the following steps: electrons or ions (depending on the sign of $\vec \Omega\cdot \vec B$) are efficiently stripped from the NS surface due to negligible cohesive energy and large temperature \citep{medin_lai2007}, and are accelerated in the strong electric field (see \S\ref{s:acceleration}). The particles gain energy $\gamma_e$ and emit curvature radiation photons with energy $E_\mathrm{CR}$. If $E_\mathrm{CR} > E_\mathrm{crit} = 10^{11}$\,eV, then they can produce electron-positron pairs in the magnetic field. These first generation particles (we use the notation of \citealt{Timokhin_2015}) are once again accelerated, and the electrons or positrons return to bombard the NS surface.

We want to investigate in detail the bottlenecks of this process, which are as follows: (a) the energy gained by electrons in the acceleration potential might not be enough to produce photons capable of electron-positron pair creation (see Section~\ref{s:acceleration}) and (b) the photon mean free path in the gap region may be much larger than the physical size of the gap itself (see Section~\ref{s:mean_free_path}).

On one hand we have a model of magnetic field evolution in the crust, which predicts the maximum curvature radius of open field lines (see \S\ref{s:actual}); on the other hand we have a constraint on this curvature radius for producing non-thermal emission (see \S\ref{s:predicted}). We assume that a NS efficiently emits non-thermal radiation only if the actual curvature radius is smaller than required for pair formation.

\subsection{Actual curvature radius of open field lines}
\label{s:actual}


Since the configuration of the poloidal magnetic field is described completely, we can integrate along the field lines. Close to the NS surface, the assumption of a vacuum magnetosphere is valid, see \cite{Gralla_2016}; in the open field line region the plasma density is small when the acceleration potential is not screened, and the assumption of vacuum also works here.


We trace magnetic field lines by numerically solving the usual system of differential equations:
\begin{align}
\frac{dr}{ds} = & \frac{B_r}{|B|} \\
\frac{d\theta}{ds} = & \frac{1}{r} \frac{B_\theta}{|B|},
\end{align} 
with the footpoints selected uniformly on the surface of neutron star given by $r_{0, i} = R_\mathrm{NS}$, $\theta_{0,i}=\pi i /500$, for the $i^{th}$ field line. We then select those which reach the light cylinder distance $R_{\mathrm{LC}} = c P / (2\pi)$, and we choose $P=0.1$\,s. To compute the curvature radius $\rho$, we follow the prescription from \cite{asseo2002} assuming flat space, since the GR corrections are small. Then
\begin{equation}
\rho (r, \theta) = \frac{1}{|(\vec b\cdot\vec \nabla) \vec b|},
\end{equation}
where $\vec{b}(r,\theta)$ is the unit vector in the direction of the local magnetic field.

\subsection{Required curvature radius for open field lines}
\label{s:predicted}

\subsubsection{Electron acceleration}
\label{s:acceleration}

We start from equation (8) from \cite{ruderman1975} for the potential difference $\Delta\,V$ between the centre of the polar cap and the edge of the negative current emission region, 
\begin{equation}
\Delta\,V \approx \Omega \frac{(r_{p-})^2}{2c} B_d^s 
\end{equation}
where $\Omega$ is the angular speed of the neutron star, $B_d^s$ is the dipole component of the magnetic field at the surface and $r_{p-}$ is the size of the polar cap. Although the gap appears to be unstable \citep{Timokhin_2010}, this vacuum acceleration potential can still be used in such studies (\citealt{Timokhin_2015, Philippov_2015}). In our case the size of the polar cap is determined by the opening angle $\theta_\mathrm{max}$ for the last open field line at the light cylinder, from 
\begin{equation}
r_{p-} = \theta_\mathrm{max} R_\mathrm{NS}.
\end{equation}
Since the light cylinder radius is $R_{\mathrm{LC}} = c/\Omega$, we obtain the potential drop
\begin{equation}
\Delta\,V \approx \frac{R_\mathrm{NS}^2 \theta_\mathrm{max}^2 B_d^s}{2R_{\mathrm{LC}}},
\end{equation}
which gives the electron Lorentz factor
\begin{equation}
\gamma = \frac{e\Delta\,V}{m_e c^2} = \frac{eR_\mathrm{NS}^2}{2R_{\mathrm{LC}} m_e c^2} (\theta_\mathrm{max}^2 B_d^s) \approx 0.586 \cdot (\theta_\mathrm{max}^2 B_d^s),
\label{e:lorentz}
\end{equation}
\noindent where $B_d^s$ has dimensions of G, and we have taken the canonical $R_{\mathrm{NS}}=10$\,km. The value for $\gamma$ should exceed $\gamma_\mathrm{crit} \approx 2\times 10^5$ for electrons to activate the cascade \citep{ruderman1975}. In this model, the acceleration potential does not depend on curvature of open magnetic field lines, and sets no conditions for the NS to exhibit pulsed emission.

The particle accelerated in the electric potential emits curvature radiation photon which can produce new electron-positron pairs in magnetic field. We consider critical values for this process in following section.

\subsubsection{Required curvature radius based on the mean free path of photons}
\label{s:mean_free_path}

The curvature radiation photon has energy \citep{ruderman1975}
\begin{equation}
E_\mathrm{CR} = \frac{3}{2} \gamma^3 \frac{\hbar c}{\rho},
\label{e:phot_en}
\end{equation}
where $\gamma$ is the Lorentz factor of the electrons from Eq.\,(\ref{e:lorentz}). 
The photon effectively produce pairs only if its mean free path
in the magnetic field is smaller than the size of the acceleration gap itself. 
The initial propagation direction aligns with the local magnetic field, but at some distance 
it starts to deviate since the magnetic field is curved. The mean free path is
determined by the strength of the orthogonal to propagation direction component
of the magnetic field ($B_\perp$). 
We assume that the radius of curvature exceeds the size of the emission zone ($h\approx 0.01R_\mathrm{NS}$\,--\,$0.1R_\mathrm{NS}$)
and a linear approximation is valid, such that
\begin{equation}
B_\perp \sim \frac{hB}{\rho}
\end{equation}
where $B$ is the strength of the total magnetic field in the emission region.\footnote{We repeatedly call the region from the surface to $1.01R_\mathrm{NS}$ at the magnetic pole as the emission region even though there might be no emission from there. } 
 We start from equation (3.1) in \cite{erber1966} and find the exponential parameter $\chi$ from
\begin{equation}
\chi = \frac{1}{2} \frac{E_\mathrm{CR}}{m_e c^2} \frac{B_\perp}{B_q} = \frac{3}{4} \frac{\hbar}{m_e c B_q} \frac{\gamma^3 h B}{\rho^2}.
\end{equation} 
Here $B_q = 4.414\times\,10^{13}$ G is the Schwinger critical magnetic field. Then, the mean free path is
\begin{equation}
l_\mathrm{CR} = 2 \frac{\hbar^2}{m_e e^2} B_q \frac{\rho}{hB} \frac{1}{T(\chi)},
\end{equation}
where we have used the approximate form for $T(\chi)$ as in \cite{erber1966}, since in our simulations the parameter $\chi$ is far from the asymptotic cases:
\begin{equation}
T(\chi) \approx 0.16 \chi^{-1} K^2_{1/3}\left(2/3\chi\right).
\end{equation}
Upon substituting numerical values we obtain
\begin{equation}
\chi = 1.3\times 10^{-25} \frac{h B}{\rho^2} (\theta_\mathrm{max}^2 B^s_d)^3,
\end{equation}
and the mean free path is
\begin{equation}
l_\mathrm{CR} = 2.8\times 10^6 \frac{\rho}{h B} \chi \frac{1}{K^2_{1/3}\left(2/3\chi\right)}.
\end{equation}

We want $l_\mathrm{CR} < h$, otherwise photons freely leave the emission region. This condition sets an upper limit for the curvature radius of the open field lines. In \S\ref{s:emis}, we solve this equation numerically and find the maximum curvature radius required for an NS to emit as a pulsar. 


\subsection{When does the pulsar shine again?}
\label{s:emis}

\begin{figure*}
\begin{minipage}{0.48\linewidth}
\includegraphics[width=84mm]{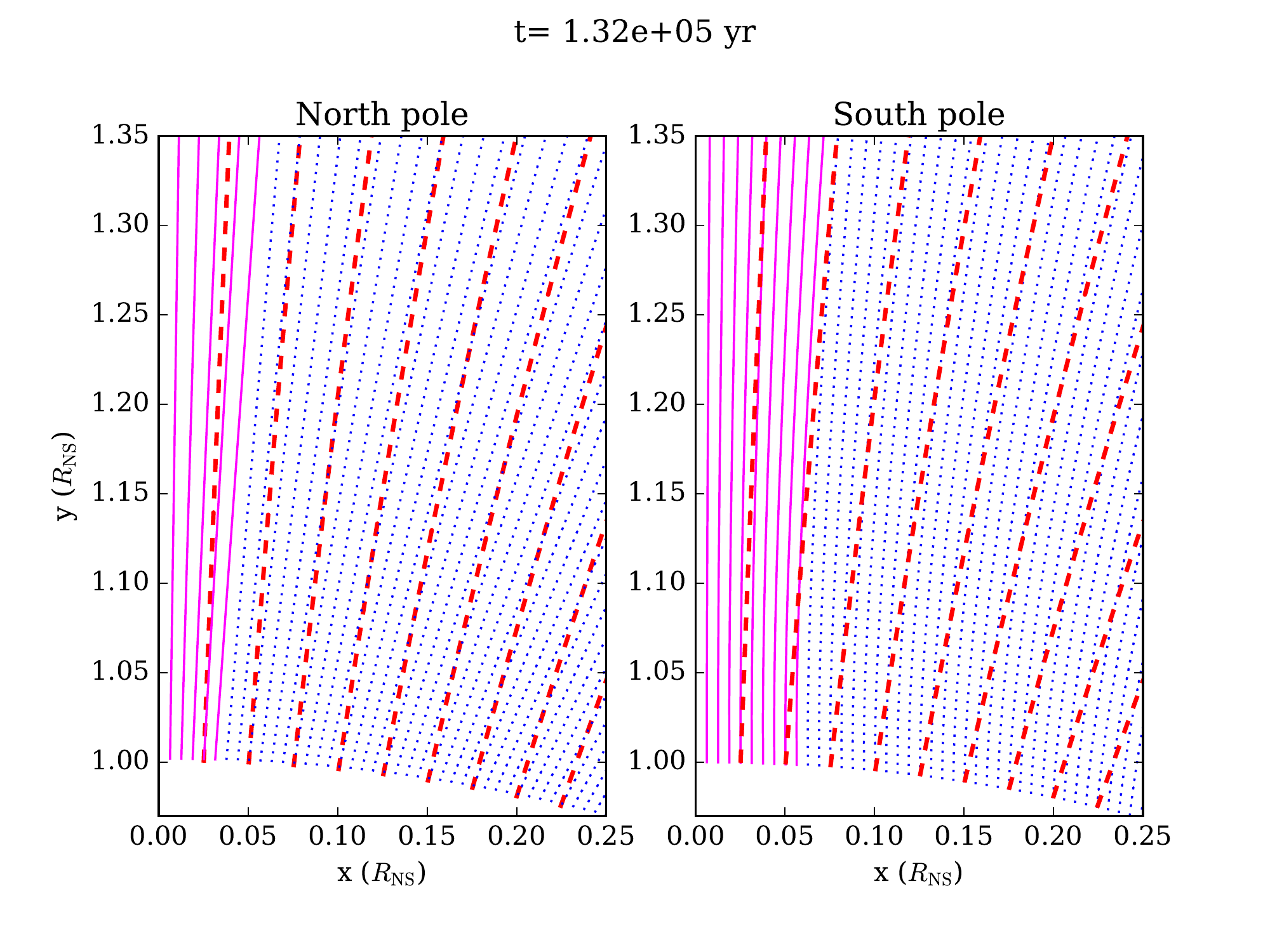}
\end{minipage}
\begin{minipage}{0.48\linewidth}
\includegraphics[width=84mm]{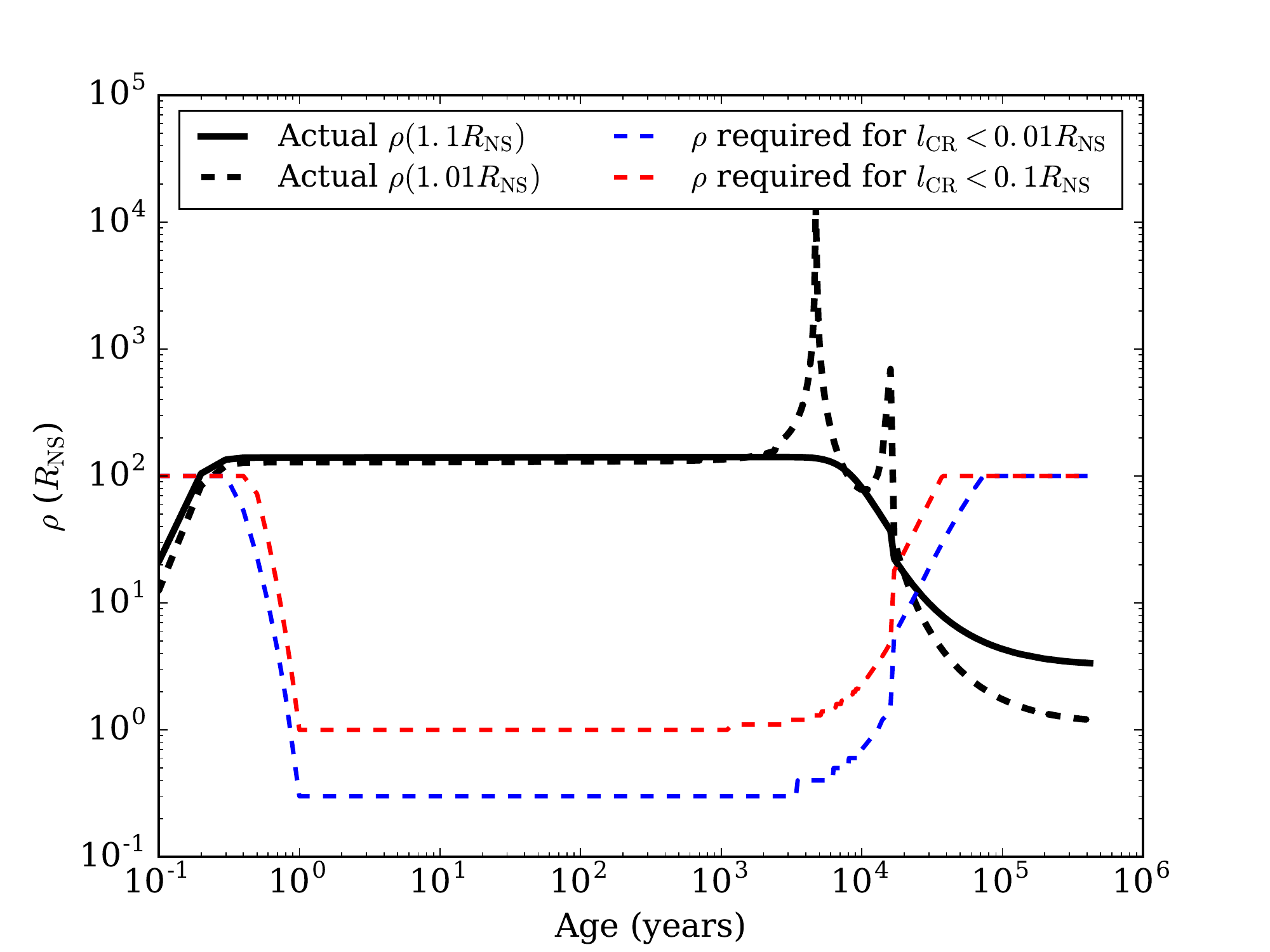}
\end{minipage}
\caption{
The structure of magnetic field lines above the NS surface (left panel) for the initial conditions consisting of dipole and $l=6$ harmonic; north and south poles both shown. The dashed red lines show purely dipolar magnetic field lines at the pole, dotted are for closed and solid are for open field lines. The right panel shows the maximum curvature radius for open field lines for the same initial configuration (black line). Other lines show theoretical predictions: blue dashed line for typical height of the emission zone $h = 0.01 R_\mathrm{NS}$; red dashed line is for larger height of the emission zone $h = 0.1 R_\mathrm{NS}$. 
}
\label{f:emis_l6}
\end{figure*}

\begin{figure*}
\begin{minipage}{0.48\linewidth}
\includegraphics[width=84mm]{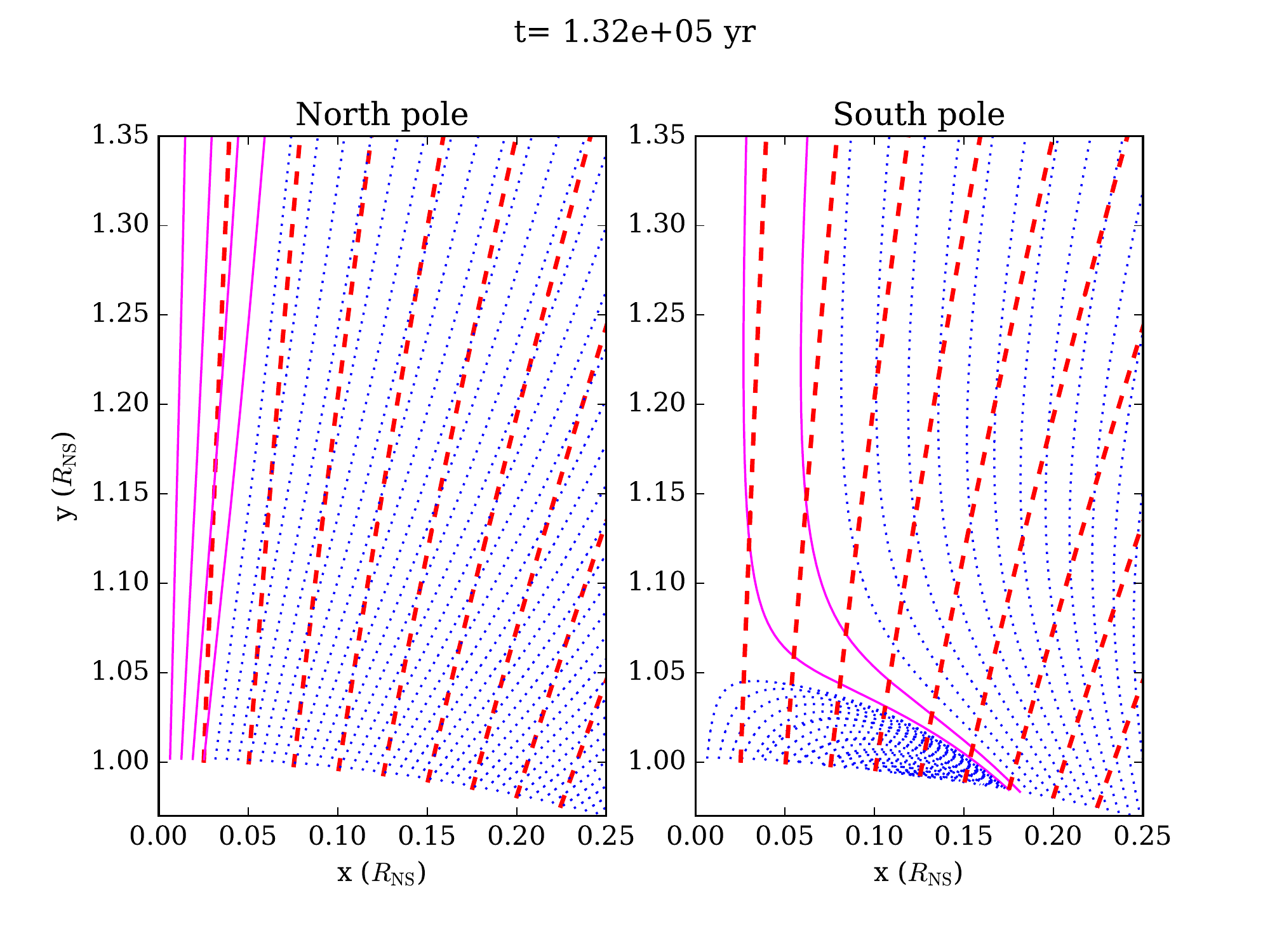}
\end{minipage}
\begin{minipage}{0.48\linewidth}
\includegraphics[width=84mm]{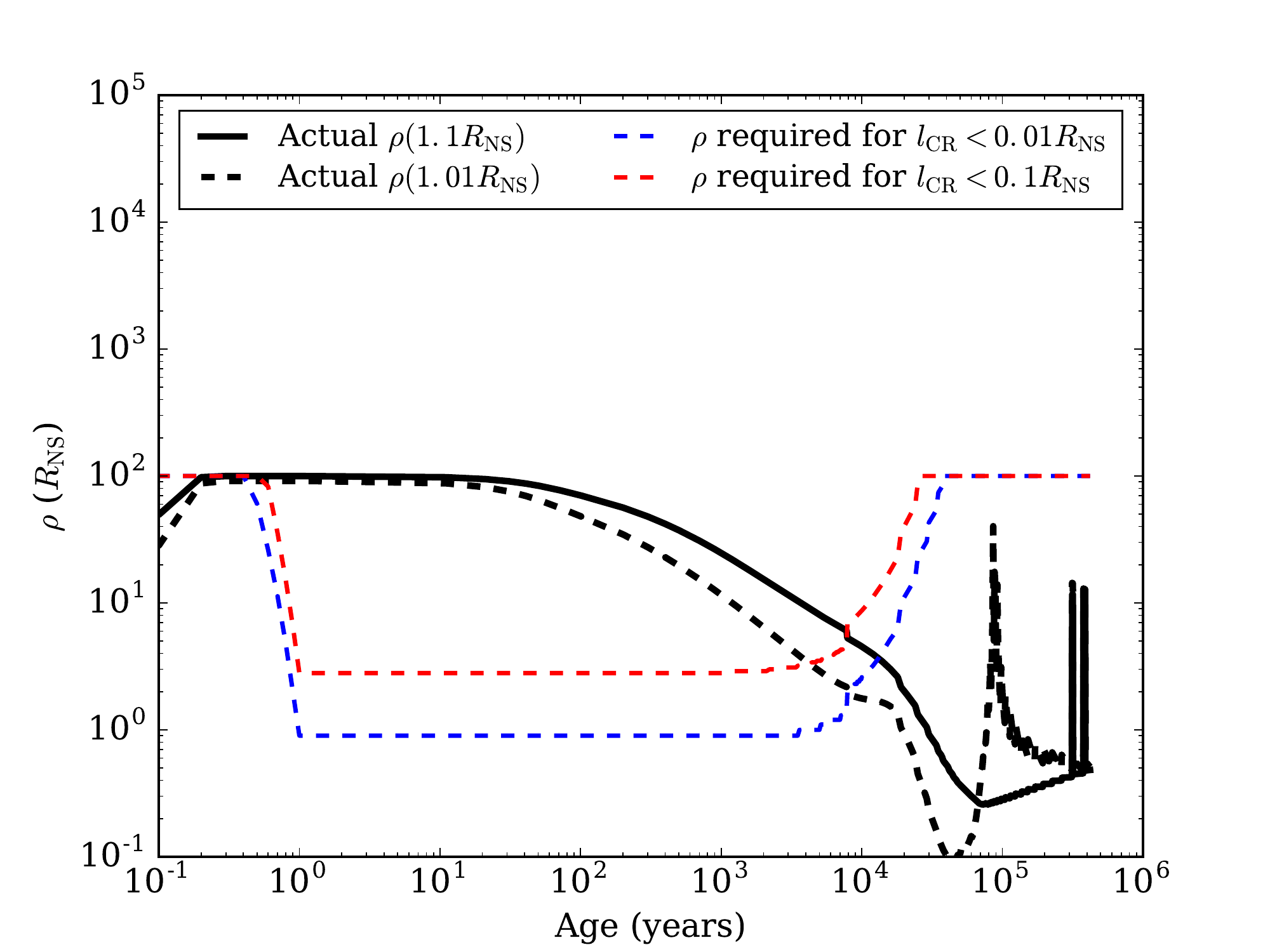}
\end{minipage}
\caption{
The structure of magnetic field lines above the NS surface (left panel) for the initial conditions consisting of dipole and $l=10$ harmonic; north and south poles both shown. The dashed red lines show purely dipolar magnetic field lines at the pole, dotted are for closed and solid are for open field lines. The right panel shows the maximum curvature radius for open field lines for the same initial configuration (black line). Other lines show theoretical predictions: blue dashed line for typical height of the emission zone $h = 0.01 R_\mathrm{NS}$; red dashed line is for larger height of the emission zone $h = 0.1 R_\mathrm{NS}$. 
}
\label{f:emis_l10}
\end{figure*}

\begin{figure*}
\begin{minipage}{0.48\linewidth}
\includegraphics[width=84mm]{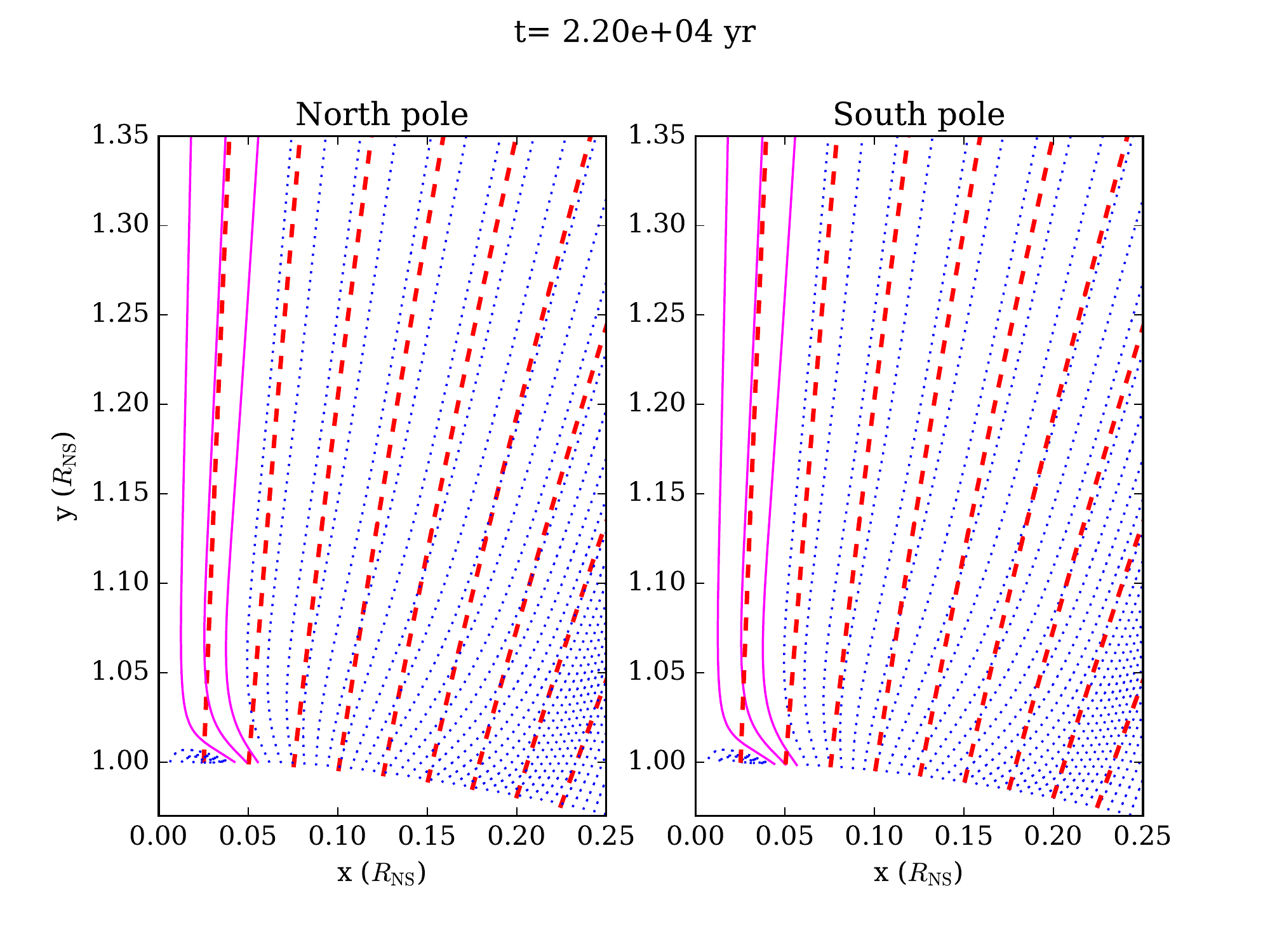}
\end{minipage}
\begin{minipage}{0.48\linewidth}
\includegraphics[width=84mm]{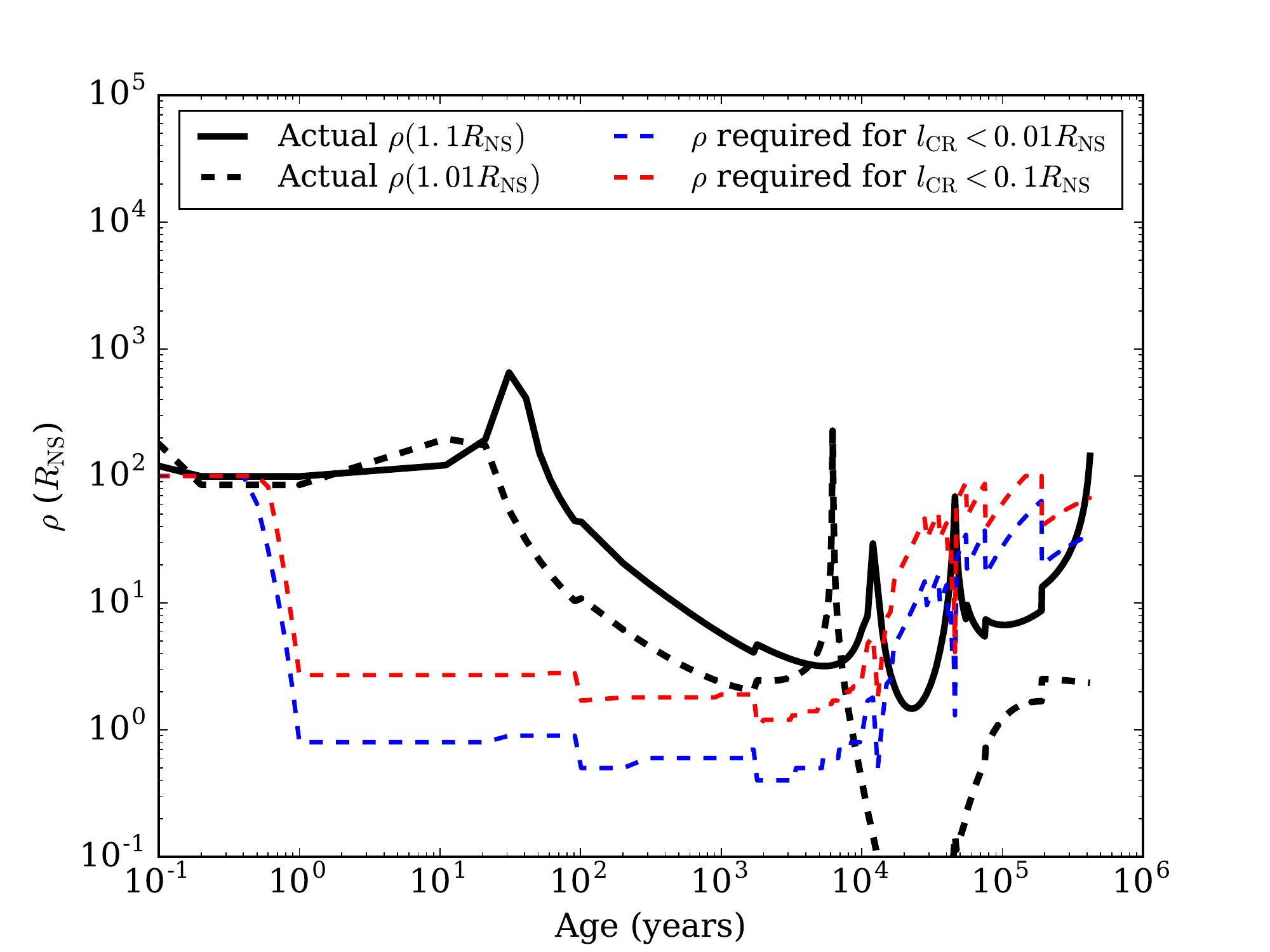}
\end{minipage}
\caption{
The structure of magnetic field lines above the NS surface (left panel; video is available at \url{http://pulsars.info/videos/l15.mp4}) for the initial conditions consisting of dipole and $l=15$ harmonic; north and south poles both shown. The dashed red lines show purely dipolar magnetic field lines at the pole, dotted are for closed and solid are for open field lines. The right panel shows the maximum curvature radius for open field lines for the same initial configuration (black line). Other lines show theoretical predictions: blue dashed line for typical height of the emission zone $h = 0.01 R_\mathrm{NS}$; red dashed line is for larger height of the emission zone $h = 0.1 R_\mathrm{NS}$. 
}
\label{f:emis_l15}
\end{figure*}

For all simulations with fall-back, we show in Figs.\,(\ref{f:emis_l6}, \ref{f:emis_l10}, \ref{f:emis_l15}) the actual (black lines) and required (blue and red lines) curvature radius of open field lines. Initially the required maximum curvature radius is larger than the actual one, which means that the NS emits as a pulsar. After the fall-back episode large harmonics are strongly suppressed and the actual curvature radius reached typical
for a pure dipole value ($\sim$100\,$R_\mathrm{NS}$). The required curvature radius for effective pairs creation decreases to extremely small
values around $0.1 R_\mathrm{NS}$ since the magnetic field is strongly suppressed and
$B_\perp$ is not developed enough.
During this period the NS does not emit non-thermal radiation. It is still visible as a source of pure thermal radiation with temperatures around $10^6$~K (see temperature labels at Fig.\,\ref{f:l2_6_10}).

Depending on the initial field configuration the actual curvature radius of open field lines starts to decrease at $\sim$10$^4$\,yrs for initial dipole and $l=6$ harmonic and $\sim$100\,yrs for initial dipole and $l=15$ harmonic. The actual and required curvatures of open field lines intersects at around a few times $10^4$\,yrs which means that the NS starts efficiently emitting non-thermal radiation.

\section{Discussion}\label{s:discussion}

\subsection{Some properties of the external magnetic field}\label{s:limitations}

The accuracy of our numerical solutions for the crustal magnetic field is restricted by the number of bins in the angular direction. From the standpoint of numerical stability, the required angular resolution must be calculated from consideration of the initial conditions. For the $l=6$ and the $l=15$ case, we use 150 and 300 angular cells, respectively. This gives a corresponding angular resolution of 1.2$^{\circ}$ and 0.6$^{\circ}$. We must also consider whether the angular resolution is sufficient from the standpoint of tracing field lines in the emission zone. For real NSs in nature, the size of the polar cap is unknown and undoubtedly varies among sources. We conclude that our chosen resolution is indeed sufficient for the following reason: the magnetospheric field satisfies our outer boundary condition, the Legendre expansion. Technically at the NS surface we perform the expansion up to order $l=200$, far higher than any order we are interested in modelling. If we were to increase the angular resolution, the dominant coefficients in the multipole expansion stay the same\,--\,i.e. the imposed initial conditions\,--\,and only the multipole coefficients corresponding to the Nyquist multipole should fluctuate. In addition, since we are not performing detailed simulations of the emission processes, but instead are only considering the magnetic topology in the region, we note that our angular resolution is at worst comparable to the polar cap size $r_{p-}$. Since our surface expansion conserves $\vec{\nabla}\cdot\vec{B}=0$ everywhere, we are free to trace the field lines with sub-grid scale resolution if desired. 



Another particularly interesting case study is the shifted dipole, which was long-discussed as a plausible explanation for the small curvature radius of open field lines in pulsars \citep{Ruderman1998}. While our numerical model would require significant modifications in order to model mis-aligned rotational and magnetic axes (see \citealt{vigano_thesis}), we could approximate such a configuration by imposing a smooth spectrum of harmonics as our initial conditions. However, looking at Fig.\,\ref{f:l2_6_10}, we can immediately conclude that the shifted dipole case would probably not survive the fall-back episode. The low-order multipoles $l=2,3,..., 6$ re-emerge less efficiently than the dipole component, while higher-order multipoles of $l=10, ..., 15$ re-emerge more efficient than the dipole component. This could lead to the destruction of the shifted dipole and would result in growth of the leading multipole, most likely the highly-structured field components.

\subsection{Limitations of Ruderman and Sutherland emission model}
\label{discussion}

We are aware that the vacuum gap model does not describe all emission process entirely. The electron work function and the ionic cohesive energy appears to be much smaller that it was expected in the time of \cite{ruderman1975}. As \cite{ml2006a, ml2006b} have shown for the case of condensed surfaces, the work function for electrons is roughly 100\,eV, and the cohesive energy is $\sim$500-700\,eV for different atmospheric compositions, in the weak magnetic field limit \footnote{If a condensate is not formed, which is possible in the case of post-CCO NSs, then the situation is uncertain. However we can assume that the cohesive energy in this case is not exceedingly large. In addition, we want to note that accretion in a fall-back episode of a significant amount of light elements can result in changes in the surface properties, which can suppress opening of the gap. If the magnetosphere is positive above the poles, then for relatively low fields and high temperatures it is much more difficult to form a gap above hydrogen than above an iron surface (see fig. 4 in \citealt{ml2007}). This can be an additional reason why PSRs do not appear immediately after the field re-emerges to the surface.
}. Then for typical surface temperatures of post-CCO NSs $\sim$0.5-1$\times 10^6$\,K, no vacuum gap can be formed because the charge can be supplied at local the Goldreich-Julian rate (\citealt{medin_lai2007}).

This problem is well-known, and has been already discussed in several papers (\citealt*{nlk1986, nkl1987}). The first solution for an extended space-charge limited flow was suggested by \cite{arons1979} for the case when large multipoles are present. The authors considered steady flow in the co-rotation frame and identified favourable field line curvature, i.e. in the direction of rotation. Moreover, consideration of the Lense-Thirring effect of frame-dragging \citep{muslimov1992} leads to the conclusion that the presence of free charge carriers at the NS surface cannot tighten the gap. The electric field grows rapidly from the surface of neutron stars and reaches a maximum at a distance of the polar cap size, and then decays. The accelerating potential is slightly different from the classical Ruderman and Sutherland potential.

However particle acceleration is not the only component of the pulsar emission mechanism. Electron-positron pairs transfer energy to photons, which then produce the next generation of pairs. Therefore it is essential to have a field topology which decreases the photon mean free path to a scale less than the size of the acceleration zone. Detailed physics of the acceleration zone has been considered in many studies, e.g. \cite{beloborodov2008}, \cite{szary_thesis}, and \cite{szary2015}. Here we intended only to look at the basic properties of multipolar re-emergence and draw corresponding conclusions about how such magnetic structure is important for non-thermal pulsar emission.

\subsection{Searches for pulsars with re-emerging magnetic fields}

\begin{figure}
\includegraphics[width=84mm]{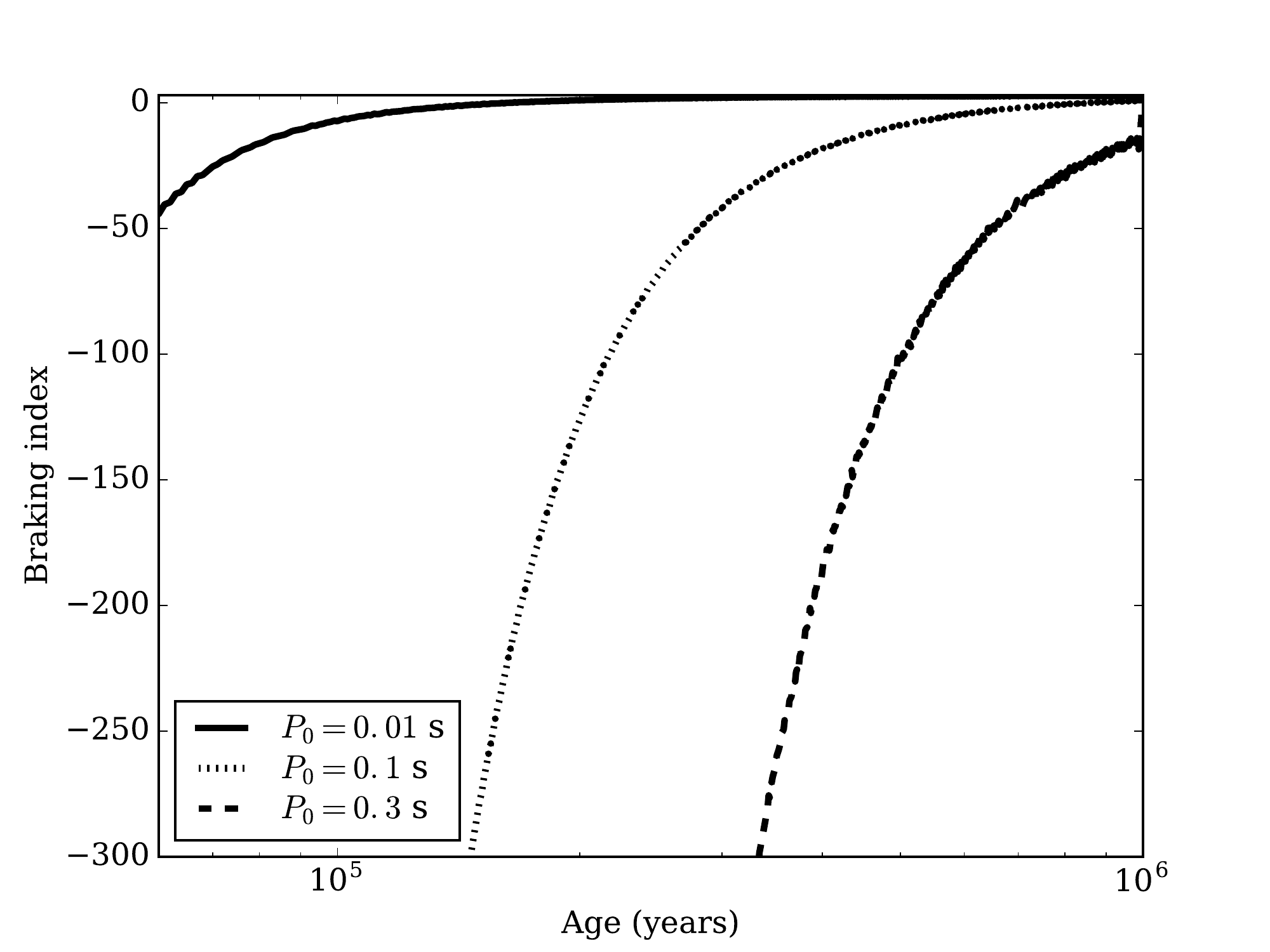}
\caption{
Evolution of braking indices based on the dipolar magnetic field component. Curves for different initial periods converge to $n=3$.
}
\label{f:n_index}
\end{figure}

Original models of magnetic re-emergence \citep{bern10, ho11, vigano2012} predicted that for ages $\gtrsim 10^4$\,yrs, the NS magnetic field would return to its initial value due to diffusion through the accreted envelope. The main observational feature of an NS at this stage should be its anomalous braking which does not correspond to the standard dipolar radiative braking. This difference can be quantified by means of the braking index, commonly written as
\begin{equation}
n = 2 - \frac{P\ddot P}{\dot P^2}.
\end{equation}
In the recent work by \cite{Ho_2015} the field re-emergence scenario was studied in detail, with the braking index formula written as 
\begin{equation}\label{e:braking}
n = 3 - 2\frac{\dot B}{B} \frac{P}{\dot P} = 3 - \frac{4}{\gamma_\mathrm{br}} \frac{\dot B}{B^3} P^2.  
\end{equation}
The $\gamma_{\mathrm{br}}$ in Eq.\,(\ref{e:braking}) is given by $\gamma_\mathrm{br} \approx 4 \pi^2 R_{\mathrm{NS}}^6 / (3c^3 I)$, with $I$ being the NS moment of inertia the NS. For typical NS parameters, we can estimate this factor as $\gamma_\mathrm{br} \approx 10^{-39}$\,G$^{-2}\,$s, and plot the braking index from the dipolar component of the total magnetic field. The results are shown in Fig.\,\ref{f:n_index}. During the re-emergence epoch, shortly after the pulsar begins emitting non thermal radiation, the braking index has extremely negative values. Such extreme values of the braking index can be impossible to measure since impulsive changes in the magnetic field could cause glitch activity according to \cite{Ho_2015}. The NS crust easily lose its torque because of coupling with crust-confined magnetic field, whereas the NS core does not. Timing noise of glitches with different magnitudes makes the second period derivative $\ddot{P}$ extremely difficult to measure. 

An alternative observational feature is that kinematically old pulsars with relatively weak fields and with braking indices $n<1$ can be associated with SNRs, and can show significant thermal emission from the NS surface. Indeed, the life-time of a SNR can be up to $10^5$\,yrs. According to standard cooling models (see, for example, \citealt{yp2004}) a CCO-like NS stays hot at least for $10^5$\,yrs, which typically corresponds to the case of light element envelopes. Light elements enhance heat transfer to the stellar surface, and so make such sources hotter and brighter at young ages \citep{vigano2013}. 

\cite{bogdanov2014} and \cite{luo2015} used two different approaches to search for such PSRs with re-emerged magnetic field. In the first paper the authors selected a sample of 8 objects with associated SNRs within the distance $d=6$~kpc. Kinematic ages of these NSs were estimated to be $\sim$20\,--\,30\,kyrs, based on their positions relative to the SNRs (and based on velocities, if available). {\it Chandra} and {\it XMM-Newton} observations put upper limits on (or directly detected) their thermal X-ray emission. The authors concluded that the selected sources do not look like evolved CCOs, and therefore their associations with SNRs might be due to chance. In \cite{luo2015} a different approach was used. Twelve PSRs were selected according to the following conditions: $B<10^{11}$\,G, $P>0.05$\,s, $z<100$\,pc, where $z$ is distance from the Galactic plane. None of them appeared to be associated with a SNR. Also, none of the sources appeared to have large thermal X-ray luminosity. Typical upper limits for temperature are $\sim$50\,--\,100\,eV. Thus, \cite{bogdanov2014} and \cite{luo2015} concluded that the proposed scenario of field re-emergence, and the corresponding appearance of PSRs is not correct. Our study demonstrates that it remains possible to bring the scenario of field evolution after fall-back into correspondence with the results of X-ray searches.
  
According to the results presented above, it takes $\gtrsim\,30$\,kyrs for small scale fields to re-emerge. Only after this time a radio pulsar can begin emitting non-thermal radiation. By this point in time, a NS can cool down below the limits obtained by \cite{bogdanov2014, luo2015}, and the SNR can become too faint to detect. We thus propose that it is necessary to extend the approaches used by \cite{bogdanov2014} and \cite{luo2015} to look for colder (and older) NSs which appear as PSRs with recently re-emerged field. Correspondingly, for the 12 sources discussed by \cite{luo2015}, it is necessary to put more stringent limits on ages and temperatures. 

For distant sources it can be difficult to probe lower temperatures due to significant interstellar absorption in the soft X-ray band. In this case, a searching strategy based on identification of a NS with a SNR is not very promising. Among the sample studied by \cite{luo2015}, more than one half are farther than 2.5\,kpc, making it difficult to detect NSs with such low temperatures. Still, for nearby sources with distances $\lesssim\,1$\,--\,2\,kpc it is possible to improve limits on the surface temperatures down to 30-40\,eV with longer \textit{Chandra} exposures (several tens of ksec), or even to detect thermal emission from them according to expectations presented in Fig.\,\ref{f:l2_6_10}. 
Correspondingly, with lower temperature limits, the limits on the ages
will shift towards larger values, up to $\sim$10$^5$ yrs.
To increase the size of the sample of objects at distances
$\lesssim1$\,--\,2~kpc we can slightly relax ranges of magnetic field and
distance from the Galactic plane, in comparison with those used by
\cite{luo2015}
This can be done, especially if for particular PSRs there are
arguments in favour of their youth (proximity to one of OB
associations, etc.).


 We have selected radio pulsars from the ATNF catalogue v1.54 (\citealt{atnf}) and cross-correlated their positions with known OB associations at distances less than $3$\,kpc from the Sun. The pulsars have been chosen according to criteria similar to those in \cite{luo2015}, but we relax the criteria somewhat. We select only pulsars within 3\,kpc from the Sun with heights above the Galactic plane of less than 200\,pc, with periods $P>0.1$\,s, and magnetic fields of $<5\times\,10^{11}$\,G. Our sample contains 37 pulsars. Note that in the ATNF, distances to most of these pulsars are estimated from the dispersion measure according to \cite{tc93}. We have used several approaches to cross-correlate the pulsar sample with known OB associations, because distances to these agglomerations of stars are not very certain.  At first, we use the catalogue of \cite{1989AJ.....98.1598B} (but also see \citealt{1995AstL...21...10M}). No significant coincidences between the list of pulsars and OB associations have been found, i.e. no pulsar detections within 100\,pc from the centre of OB associations. We then take into account that distances to OB associations might be 20\% smaller (\citealt*{2001AstL...27...58D}, \citealt{2009MNRAS.400..518M}). Taking this into account, we find two pairs of pulsar-OB association: J1107-5907 and J1154-6250.

The first pulsar J1107-5907 is close to the Car OB2 association with an age of $\sim$4\,Myr (\citealt{2010MNRAS.402.2369T}). The characteristic age of the pulsar is very large, $\sim$4.5$\times10^8$\,yrs. The young age of the association and that the pulsar is nearby the association indicates that the pulsar might itself be very young, as only the most massive NS progenitors are expected to explode. Note that this object has also been studied by \cite{luo2015}. However, they assumed a distance of 1.3\,kpc, estimated according to \cite{ne2001}. With this distance the source does not fall close to the Car OB2. But new estimates based on \cite{M2} show that the distance is about 1.94\,kpc, which is in good correspondence with the ATNF value (1.81\,kpc). With this improved distance estimate, the pulsar is close to the OB association, and could thus be related to it when taking into account the uncertainties in distance. The temperature limit given by \cite{luo2015} is therefore modified, as the source is further away by factor $\sim$1.5. Since \cite{luo2015} already estimated the column density $n_{\mathrm{H}}$ from DM using the standard relation from \cite*{2013ApJ...768...64H}, we must only take into account changes in distance. Then the updated temperature limit is $\sim$67\,eV. This is not strict enough to draw clear conclusions, but is notably higher than in \cite{luo2015}. This temperature of $7.8\times 10^5$\,K corresponds to the rising part of the curve in Fig.\,\ref{f:l2_6_10}. However, 55\,eV\,--\,$6.4\times 10^5$~K\,--\,is already behind the rising part. We suggest that deeper observations of this source are necessary.

The second pulsar is J1154-6250, which can be related to the Cru OB1 association with the age 5-7~Myrs (\citealt{2010MNRAS.402.2369T}). There are no available temperature estimates for this object. With the distance calculated according to \cite{M2}, the source still remains close to the association. The characteristic age of this pulsar is $\sim$8$\times 10^6$\,yrs, inconsistent with the age of the association. Thus this source was possibly born with a spin period close to the present day value.
 
We have also used the list of 25 OB association from \cite{2001AstL...27...58D} for which parallax distances are estimated. These distances are not considered to be precise enough (\citealt{2009MNRAS.400..518M, 2001AstL...27...58D}), and in addition we have not found any pulsar from our list within 100\,pc of any of these associations, so we do not comment on it further.

Potentially, a full account of the uncertainties in pulsar distances and OB associations, of different combinations of selection parameters of pulsars, and usage of larger lists of associations could result in new cases of pulsar-association pairs. Such cases must be studied in detail, although that analysis is beyond the scope of this paper. For the purposes of this work, we simply point out examples of pulsars which could in fact be objects with re-emerging magnetic field.

\section{Conclusions}\label{s:conclusions}

We have studied, for the first time, the evolution of high-order multipolar fields ($l>10$) in a self-consistent, 2D magneto-thermal framework, while imposing a short accretion epoch as an initial condition. 
We confirm re-emergence time scales of $\sim$10$^5$\,yrs for magnetic field buried by fall-back with accreted mass $10^{-3} M_\odot$ just after NS birth \citep{vigano2012}. For our relatively weak field strengths, we find that harmonics up to $l=15$ efficiently re-emerge on time scales comparable to\,--\,or shorter than\,--\,the corresponding time scale for the dipolar magnetic field component. We also report that for high-order multipoles the toroidal-poloidal interaction plays an essential role in accelerating magnetic re-emergence.


We have implemented the full \cite{ruderman1975} formalism for strongly non-dipolar surface fields, and have confirmed that the observable field following the fall-back episode (first $\sim$10$^4$\,yrs) is extremely weak and also purely dipolar. Such conditions prevent effective conversion of photons into electron-positron pairs. The re-emergence of large multipoles at $3\times 10^4 - 5\times 10^4$\,yrs decreases the curvature radius in the emission zone. This activates non-thermal emission, and thus the neutron star manifests as a pulsar. The surface temperature at these times is about $7\times 10^5$\,K which prevents effective detection. 

Although earlier searches for pulsars with re-emerging fields have revealed no plausible candidates \citep{bogdanov2014, luo2015} we argue that our scenario is still valid, especially because the distances to the 
pulsars are highly uncertain and some candidates might be hotter than that predicted by \cite{luo2015}. We develop a new criterion for such searches, namely the small projected distance from an OB association. 
We have found two candidates, J1107-5907 and J1154-6250, which based on our simulations could be young pulsars currently experiencing magnetic re-emergence.

\section*{Acknowledgments}
AI would like to thank NOVA PhD funding. JGE acknowledges support from a Vidi grant in the NWO framework (PI: Nanda Rea). SBP is supported by the RFBR grant 14-02-00657. AI would like to thank Alexander Phillipov, Andrzej Szary, Jason Hessels, Sterl Phinney, Serena Repetto, Cameron Van Eck, and Dmitriy Barsukov for myriad constructive discussions. SBP thanks Anna Mel'nik for consultations on OB associations. The authors also thank the anonymous referee for constructive comments which improved the quality of this manuscript.

\bibliographystyle{mnras}
\bibliography{ref_list}

\begin{thebibliography}{}
\makeatletter
\relax
\def\mn@urlcharsother{\let\do\@makeother \do\$\do\&\do\#\do\^\do\_\do\%\do\~}
\def\mn@doi{\begingroup\mn@urlcharsother \@ifnextchar [ {\mn@doi@}
  {\mn@doi@[]}}
\def\mn@doi@[#1]#2{\def\@tempa{#1}\ifx\@tempa\@empty \href
  {http://dx.doi.org/#2} {doi:#2}\else \href {http://dx.doi.org/#2} {#1}\fi
  \endgroup}
\def\mn@eprint#1#2{\mn@eprint@#1:#2::\@nil}
\def\mn@eprint@arXiv#1{\href {http://arxiv.org/abs/#1} {{\tt arXiv:#1}}}
\def\mn@eprint@dblp#1{\href {http://dblp.uni-trier.de/rec/bibtex/#1.xml}
  {dblp:#1}}
\def\mn@eprint@#1:#2:#3:#4\@nil{\def\@tempa {#1}\def\@tempb {#2}\def\@tempc
  {#3}\ifx \@tempc \@empty \let \@tempc \@tempb \let \@tempb \@tempa \fi \ifx
  \@tempb \@empty \def\@tempb {arXiv}\fi \@ifundefined
  {mn@eprint@\@tempb}{\@tempb:\@tempc}{\expandafter \expandafter \csname
  mn@eprint@\@tempb\endcsname \expandafter{\@tempc}}}

\bibitem[\protect\citeauthoryear{{Aguilera}, {Pons}  \& {Miralles}}{{Aguilera}
  et~al.}{2008}]{Aguilera2008}
{Aguilera} D.~N.,  {Pons} J.~A.,   {Miralles} J.~A.,  2008, \mn@doi [\aap]
  {10.1051/0004-6361:20078786}, \href
  {http://adsabs.harvard.edu/abs/2008A%26A...486..255A} {486, 255}

\bibitem[\protect\citeauthoryear{{Arons} \& {Scharlemann}}{{Arons} \&
  {Scharlemann}}{1979}]{arons1979}
{Arons} J.,  {Scharlemann} E.~T.,  1979, \mn@doi [\apj] {10.1086/157250}, \href
  {http://adsabs.harvard.edu/abs/1979ApJ...231..854A} {231, 854}

\bibitem[\protect\citeauthoryear{{Asseo} \& {Khechinashvili}}{{Asseo} \&
  {Khechinashvili}}{2002}]{asseo2002}
{Asseo} E.,  {Khechinashvili} D.,  2002, \mn@doi [\mnras]
  {10.1046/j.1365-8711.2002.05481.x}, \href
  {http://adsabs.harvard.edu/abs/2002MNRAS.334..743A} {334, 743}

\bibitem[\protect\citeauthoryear{{Baym}, {Bethe}  \& {Pethick}}{{Baym}
  et~al.}{1971}]{Baym1971}
{Baym} G.,  {Bethe} H.~A.,   {Pethick} C.~J.,  1971, \mn@doi [Nuclear Phys. A]
  {10.1016/0375-9474(71)90281-8}, \href
  {http://adsabs.harvard.edu/abs/1971NuPhA.175..225B} {175, 225}

\bibitem[\protect\citeauthoryear{{Beloborodov}}{{Beloborodov}}{2008}]{beloborodov2008}
{Beloborodov} A.~M.,  2008, \mn@doi [\apjl] {10.1086/590079}, \href
  {http://adsabs.harvard.edu/abs/2008ApJ...683L..41B} {683, L41}

\bibitem[\protect\citeauthoryear{{Bernal} \& {Fraija}}{{Bernal} \&
  {Fraija}}{2016}]{bernal2016}
{Bernal} C.~G.,  {Fraija} N.,  2016, preprint, \href
  {http://adsabs.harvard.edu/abs/2016arXiv160705652B} {} (\mn@eprint {arXiv}
  {1607.05652})

\bibitem[\protect\citeauthoryear{{Bernal}, {Lee}  \& {Page}}{{Bernal}
  et~al.}{2010}]{bern10}
{Bernal} C.~G.,  {Lee} W.~H.,   {Page} D.,  2010, \rmxaa, \href
  {http://adsabs.harvard.edu/abs/2010RMxAA..46..309B} {46, 309}

\bibitem[\protect\citeauthoryear{{Bernal}, {Page}  \& {Lee}}{{Bernal}
  et~al.}{2013}]{bern12}
{Bernal} C.~G.,  {Page} D.,   {Lee} W.~H.,  2013, \mn@doi [\apj]
  {10.1088/0004-637X/770/2/106}, \href
  {http://adsabs.harvard.edu/abs/2013ApJ...770..106B} {770, 106}

\bibitem[\protect\citeauthoryear{{Bhattacharya}, {Wijers}, {Hartman}  \&
  {Verbunt}}{{Bhattacharya} et~al.}{1992}]{bhattacharya1992}
{Bhattacharya} D.,  {Wijers} R.~A.~M.~J.,  {Hartman} J.~W.,   {Verbunt} F.,
  1992, \aap, \href {http://adsabs.harvard.edu/abs/1992A%26A...254..198B} {254,
  198}

\bibitem[\protect\citeauthoryear{{Blaha} \& {Humphreys}}{{Blaha} \&
  {Humphreys}}{1989}]{1989AJ.....98.1598B}
{Blaha} C.,  {Humphreys} R.~M.,  1989, \mn@doi [\aj] {10.1086/115244}, \href
  {http://adsabs.harvard.edu/abs/1989AJ.....98.1598B} {98, 1598}

\bibitem[\protect\citeauthoryear{{Bogdanov}, {Ng}  \& {Kaspi}}{{Bogdanov}
  et~al.}{2014}]{bogdanov2014}
{Bogdanov} S.,  {Ng} C.-Y.,   {Kaspi} V.~M.,  2014, \mn@doi [\apjl]
  {10.1088/2041-8205/792/2/L36}, \href
  {http://adsabs.harvard.edu/abs/2014ApJ...792L..36B} {792, L36}

\bibitem[\protect\citeauthoryear{{Chashkina} \& {Popov}}{{Chashkina} \&
  {Popov}}{2012}]{cp2012}
{Chashkina} A.,  {Popov} S.~B.,  2012, \mn@doi [\na]
  {10.1016/j.newast.2012.01.004}, \href
  {http://adsabs.harvard.edu/abs/2012NewA...17..594C} {17, 594}

\bibitem[\protect\citeauthoryear{{Chevalier}}{{Chevalier}}{1989}]{chevalier}
{Chevalier} R.~A.,  1989, \mn@doi [\apj] {10.1086/168066}, \href
  {http://adsabs.harvard.edu/abs/1989ApJ...346..847C} {346, 847}

\bibitem[\protect\citeauthoryear{{Cordes} \& {Lazio}}{{Cordes} \&
  {Lazio}}{2002}]{ne2001}
{Cordes} J.~M.,  {Lazio} T.~J.~W.,  2002, preprint, \href
  {http://adsabs.harvard.edu/abs/2002astro.ph..7156C} {} (\mn@eprint {arXiv}
  {0207156})

\bibitem[\protect\citeauthoryear{{Cumming}, {Arras}  \& {Zweibel}}{{Cumming}
  et~al.}{2004}]{cumming2004}
{Cumming} A.,  {Arras} P.,   {Zweibel} E.,  2004, \mn@doi [\apj]
  {10.1086/421324}, \href {http://adsabs.harvard.edu/abs/2004ApJ...609..999C}
  {609, 999}

\bibitem[\protect\citeauthoryear{{Dambis}, {Mel'nik}  \& {Rastorguev}}{{Dambis}
  et~al.}{2001}]{2001AstL...27...58D}
{Dambis} A.~K.,  {Mel'nik} A.~M.,   {Rastorguev} A.~S.,  2001, \mn@doi [Astron.
  Letters] {10.1134/1.1336863}, \href
  {http://adsabs.harvard.edu/abs/2001AstL...27...58D} {27, 58}

\bibitem[\protect\citeauthoryear{{Douchin} \& {Haensel}}{{Douchin} \&
  {Haensel}}{2001}]{Douchin2001}
{Douchin} F.,  {Haensel} P.,  2001, \mn@doi [\aap]
  {10.1051/0004-6361:20011402}, \href
  {http://adsabs.harvard.edu/abs/2001A%26A...380..151D} {380, 151}

\bibitem[\protect\citeauthoryear{{Elfritz}, {Pons}, {Rea}, {Glampedakis}  \&
  {Vigan{\`o}}}{{Elfritz} et~al.}{2016}]{Elfritz2016}
{Elfritz} J.~G.,  {Pons} J.~A.,  {Rea} N.,  {Glampedakis} K.,   {Vigan{\`o}}
  D.,  2016, \mn@doi [\mnras] {10.1093/mnras/stv2963}, \href
  {http://adsabs.harvard.edu/abs/2016MNRAS.456.4461E} {456, 4461}

\bibitem[\protect\citeauthoryear{{Erber}}{{Erber}}{1966}]{erber1966}
{Erber} T.,  1966, \mn@doi [Rev. of Modern Phys.] {10.1103/RevModPhys.38.626},
  \href {http://adsabs.harvard.edu/abs/1966RvMP...38..626E} {38, 626}

\bibitem[\protect\citeauthoryear{{Geppert}, {Page}  \& {Zannias}}{{Geppert}
  et~al.}{1999}]{hidden}
{Geppert} U.,  {Page} D.,   {Zannias} T.,  1999, \aap, \href
  {http://adsabs.harvard.edu/abs/1999A%26A...345..847G} {345, 847}

\bibitem[\protect\citeauthoryear{{Gonthier}, {Ouellette}, {Berrier}, {O'Brien}
  \& {Harding}}{{Gonthier} et~al.}{2002}]{gonthier2002}
{Gonthier} P.~L.,  {Ouellette} M.~S.,  {Berrier} J.,  {O'Brien} S.,   {Harding}
  A.~K.,  2002, \mn@doi [\apj] {10.1086/324535}, \href
  {http://adsabs.harvard.edu/abs/2002ApJ...565..482G} {565, 482}

\bibitem[\protect\citeauthoryear{{Gotthelf}, {Halpern}  \& {Alford}}{{Gotthelf}
  et~al.}{2013}]{gotthelf2013}
{Gotthelf} E.~V.,  {Halpern} J.~P.,   {Alford} J.,  2013, \mn@doi [\apj]
  {10.1088/0004-637X/765/1/58}, \href
  {http://adsabs.harvard.edu/abs/2013ApJ...765...58G} {765, 58}

\bibitem[\protect\citeauthoryear{{Gourgouliatos} \& {Cumming}}{{Gourgouliatos}
  \& {Cumming}}{2014}]{Gourgouliatos2014}
{Gourgouliatos} K.~N.,  {Cumming} A.,  2014, \mn@doi [Phys. Rev. Letters]
  {10.1103/PhysRevLett.112.171101}, \href
  {http://adsabs.harvard.edu/abs/2014PhRvL.112q1101G} {112, 171101}

\bibitem[\protect\citeauthoryear{{Gralla}, {Lupsasca}  \& {Philippov}}{{Gralla}
  et~al.}{2016}]{Gralla_2016}
{Gralla} S.~E.,  {Lupsasca} A.,   {Philippov} A.,  2016, preprint, \href
  {http://adsabs.harvard.edu/abs/2016arXiv160404625G} {} (\mn@eprint {arXiv}
  {1604.04625})

\bibitem[\protect\citeauthoryear{{Gull{\'o}n}, {Miralles}, {Vigan{\`o}}  \&
  {Pons}}{{Gull{\'o}n} et~al.}{2014}]{gullon2014}
{Gull{\'o}n} M.,  {Miralles} J.~A.,  {Vigan{\`o}} D.,   {Pons} J.~A.,  2014,
  \mn@doi [\mnras] {10.1093/mnras/stu1253}, \href
  {http://adsabs.harvard.edu/abs/2014MNRAS.443.1891G} {443, 1891}

\bibitem[\protect\citeauthoryear{{Gull{\'o}n}, {Pons}, {Miralles},
  {Vigan{\`o}}, {Rea}  \& {Perna}}{{Gull{\'o}n} et~al.}{2015}]{gullon2015}
{Gull{\'o}n} M.,  {Pons} J.~A.,  {Miralles} J.~A.,  {Vigan{\`o}} D.,  {Rea} N.,
    {Perna} R.,  2015, \mn@doi [\mnras] {10.1093/mnras/stv1644}, \href
  {http://adsabs.harvard.edu/abs/2015MNRAS.454..615G} {454, 615}

\bibitem[\protect\citeauthoryear{{Halpern} \& {Gotthelf}}{{Halpern} \&
  {Gotthelf}}{2010}]{halpern2010}
{Halpern} J.~P.,  {Gotthelf} E.~V.,  2010, \mn@doi [\apj]
  {10.1088/0004-637X/709/1/436}, \href
  {http://adsabs.harvard.edu/abs/2010ApJ...709..436H} {709, 436}

\bibitem[\protect\citeauthoryear{{Harding}}{{Harding}}{2013}]{Harding2013}
{Harding} A.~K.,  2013, \mn@doi [Frontiers of Phys.]
  {10.1007/s11467-013-0285-0}, \href
  {http://adsabs.harvard.edu/abs/2013FrPhy...8..679H} {8, 679}

\bibitem[\protect\citeauthoryear{{He}, {Ng}  \& {Kaspi}}{{He}
  et~al.}{2013}]{2013ApJ...768...64H}
{He} C.,  {Ng} C.-Y.,   {Kaspi} V.~M.,  2013, \mn@doi [\apj]
  {10.1088/0004-637X/768/1/64}, \href
  {http://adsabs.harvard.edu/abs/2013ApJ...768...64H} {768, 64}

\bibitem[\protect\citeauthoryear{{Ho}}{{Ho}}{2011}]{ho11}
{Ho} W.~C.~G.,  2011, \mn@doi [\mnras] {10.1111/j.1365-2966.2011.18576.x},
  \href {http://adsabs.harvard.edu/abs/2011MNRAS.414.2567H} {414, 2567}

\bibitem[\protect\citeauthoryear{{Ho}}{{Ho}}{2015}]{Ho_2015}
{Ho} W.~C.~G.,  2015, \mn@doi [\mnras] {10.1093/mnras/stv1339}, \href
  {http://adsabs.harvard.edu/abs/2015MNRAS.452..845H} {452, 845}

\bibitem[\protect\citeauthoryear{{Igoshev} \& {Popov}}{{Igoshev} \&
  {Popov}}{2013}]{igoshev2013}
{Igoshev} A.~P.,  {Popov} S.~B.,  2013, \mn@doi [\mnras]
  {10.1093/mnras/stt519}, \href
  {http://adsabs.harvard.edu/abs/2013MNRAS.432..967I} {432, 967}

\bibitem[\protect\citeauthoryear{{Igoshev}, {Popov}  \& {Turolla}}{{Igoshev}
  et~al.}{2014}]{ipt2014}
{Igoshev} A.~P.,  {Popov} S.~B.,   {Turolla} R.,  2014, \mn@doi [Astronomische
  Nachrichten] {10.1002/asna.201312029}, \href
  {http://adsabs.harvard.edu/abs/2014AN....335..262I} {335, 262}

\bibitem[\protect\citeauthoryear{{Kaspi}}{{Kaspi}}{2010}]{guns}
{Kaspi} V.~M.,  2010, \mn@doi [Proc. of the Natl. Acad. of Sci.]
  {10.1073/pnas.1000812107}, \href
  {http://adsabs.harvard.edu/abs/2010PNAS..107.7147K} {107, 7147}

\bibitem[\protect\citeauthoryear{{Luo}, {Ng}, {Ho}, {Bogdanovc}, {Kaspi}  \&
  {He}}{{Luo} et~al.}{2015}]{luo2015}
{Luo} J.,  {Ng} C.-Y.,  {Ho} W.~C.~G.,  {Bogdanovc} S.,  {Kaspi} V.~M.,   {He}
  C.,  2015, \mn@doi [\apj] {10.1088/0004-637X/808/2/130}, \href
  {http://adsabs.harvard.edu/abs/2015ApJ...808..130L} {808, 130}

\bibitem[\protect\citeauthoryear{{Manchester}, {Hobbs}, {Teoh}  \&
  {Hobbs}}{{Manchester} et~al.}{2005}]{atnf}
{Manchester} R.~N.,  {Hobbs} G.~B.,  {Teoh} A.,   {Hobbs} M.,  2005, \mn@doi
  [\aj] {10.1086/428488}, \href
  {http://adsabs.harvard.edu/abs/2005AJ....129.1993M} {129, 1993}

\bibitem[\protect\citeauthoryear{{Medin} \& {Lai}}{{Medin} \&
  {Lai}}{2006a}]{ml2006a}
{Medin} Z.,  {Lai} D.,  2006a, \mn@doi [\pra] {10.1103/PhysRevA.74.062507},
  \href {http://adsabs.harvard.edu/abs/2006PhRvA..74f2507M} {74, 062507}

\bibitem[\protect\citeauthoryear{{Medin} \& {Lai}}{{Medin} \&
  {Lai}}{2006b}]{ml2006b}
{Medin} Z.,  {Lai} D.,  2006b, \mn@doi [\pra] {10.1103/PhysRevA.74.062508},
  \href {http://adsabs.harvard.edu/abs/2006PhRvA..74f2508M} {74, 062508}

\bibitem[\protect\citeauthoryear{{Medin} \& {Lai}}{{Medin} \&
  {Lai}}{2007a}]{ml2007}
{Medin} Z.,  {Lai} D.,  2007a, \mn@doi [Advances in Space Res.]
  {10.1016/j.asr.2007.01.036}, \href
  {http://adsabs.harvard.edu/abs/2007AdSpR..40.1466M} {40, 1466}

\bibitem[\protect\citeauthoryear{{Medin} \& {Lai}}{{Medin} \&
  {Lai}}{2007b}]{medin_lai2007}
{Medin} Z.,  {Lai} D.,  2007b, \mn@doi [\mnras]
  {10.1111/j.1365-2966.2007.12492.x}, \href
  {http://adsabs.harvard.edu/abs/2007MNRAS.382.1833M} {382, 1833}

\bibitem[\protect\citeauthoryear{{Mel'nik} \& {Dambis}}{{Mel'nik} \&
  {Dambis}}{2009}]{2009MNRAS.400..518M}
{Mel'nik} A.~M.,  {Dambis} A.~K.,  2009, \mn@doi [\mnras]
  {10.1111/j.1365-2966.2009.15484.x}, \href
  {http://adsabs.harvard.edu/abs/2009MNRAS.400..518M} {400, 518}

\bibitem[\protect\citeauthoryear{{Mel'nik} \& {Efremov}}{{Mel'nik} \&
  {Efremov}}{1995}]{1995AstL...21...10M}
{Mel'nik} A.~M.,  {Efremov} Y.~N.,  1995, Astron. Letters, \href
  {http://adsabs.harvard.edu/abs/1995AstL...21...10M} {21, 10}

\bibitem[\protect\citeauthoryear{{Muslimov} \& {Tsygan}}{{Muslimov} \&
  {Tsygan}}{1992}]{muslimov1992}
{Muslimov} A.~G.,  {Tsygan} A.~I.,  1992, \mn@doi [\mnras]
  {10.1093/mnras/255.1.61}, \href
  {http://adsabs.harvard.edu/abs/1992MNRAS.255...61M} {255, 61}

\bibitem[\protect\citeauthoryear{{Neuhauser}, {Langanke}  \&
  {Koonin}}{{Neuhauser} et~al.}{1986}]{nlk1986}
{Neuhauser} D.,  {Langanke} K.,   {Koonin} S.~E.,  1986, \mn@doi [\pra]
  {10.1103/PhysRevA.33.2084}, \href
  {http://adsabs.harvard.edu/abs/1986PhRvA..33.2084N} {33, 2084}

\bibitem[\protect\citeauthoryear{{Neuhauser}, {Koonin}  \&
  {Langanke}}{{Neuhauser} et~al.}{1987}]{nkl1987}
{Neuhauser} D.,  {Koonin} S.~E.,   {Langanke} K.,  1987, \mn@doi [\pra]
  {10.1103/PhysRevA.36.4163}, \href
  {http://adsabs.harvard.edu/abs/1987PhRvA..36.4163N} {36, 4163}

\bibitem[\protect\citeauthoryear{{Philippov}, {Cerutti}, {Tchekhovskoy}  \&
  {Spitkovsky}}{{Philippov} et~al.}{2015}]{Philippov_2015}
{Philippov} A.~A.,  {Cerutti} B.,  {Tchekhovskoy} A.,   {Spitkovsky} A.,  2015,
  \mn@doi [\apjl] {10.1088/2041-8205/815/2/L19}, \href
  {http://adsabs.harvard.edu/abs/2015ApJ...815L..19P} {815, L19}

\bibitem[\protect\citeauthoryear{{Pons} \& {Geppert}}{{Pons} \&
  {Geppert}}{2007}]{pons2007}
{Pons} J.~A.,  {Geppert} U.,  2007, \mn@doi [\aap]
  {10.1051/0004-6361:20077456}, \href
  {http://adsabs.harvard.edu/abs/2007A%26A...470..303P} {470, 303}

\bibitem[\protect\citeauthoryear{{Pons}, {Miralles}  \& {Geppert}}{{Pons}
  et~al.}{2009}]{Pons2009}
{Pons} J.~A.,  {Miralles} J.~A.,   {Geppert} U.,  2009, \mn@doi [\aap]
  {10.1051/0004-6361:200811229}, \href
  {http://adsabs.harvard.edu/abs/2009A%26A...496..207P} {496, 207}

\bibitem[\protect\citeauthoryear{{Pons}, {Vigan{\`o}}  \& {Geppert}}{{Pons}
  et~al.}{2012}]{geppert}
{Pons} J.~A.,  {Vigan{\`o}} D.,   {Geppert} U.,  2012, \mn@doi [\aap]
  {10.1051/0004-6361/201220091}, \href
  {http://adsabs.harvard.edu/abs/2012A%26A...547A...9P} {547, A9}

\bibitem[\protect\citeauthoryear{{Popov} \& {Turolla}}{{Popov} \&
  {Turolla}}{2012}]{popov2012}
{Popov} S.~B.,  {Turolla} R.,  2012, \mn@doi [\apss]
  {10.1007/s10509-012-1100-z}, \href
  {http://adsabs.harvard.edu/abs/2012Ap%26SS.341..457P} {341, 457}

\bibitem[\protect\citeauthoryear{{Popov}, {Pons}, {Miralles}, {Boldin}  \&
  {Posselt}}{{Popov} et~al.}{2010}]{popov}
{Popov} S.~B.,  {Pons} J.~A.,  {Miralles} J.~A.,  {Boldin} P.~A.,   {Posselt}
  B.,  2010, \mn@doi [\mnras] {10.1111/j.1365-2966.2009.15850.x}, \href
  {http://adsabs.harvard.edu/abs/2010MNRAS.401.2675P} {401, 2675}

\bibitem[\protect\citeauthoryear{{Popov}, {Kaurov}  \& {Kaminker}}{{Popov}
  et~al.}{2015}]{rcw103}
{Popov} S.~B.,  {Kaurov} A.~A.,   {Kaminker} A.~D.,  2015, \mn@doi [\pasa]
  {10.1017/pasa.2015.18}, \href
  {http://adsabs.harvard.edu/abs/2015PASA...32...18P} {32, e018}

\bibitem[\protect\citeauthoryear{{Rea}, {Pons}, {Torres}  \& {Turolla}}{{Rea}
  et~al.}{2012}]{rea2012}
{Rea} N.,  {Pons} J.~A.,  {Torres} D.~F.,   {Turolla} R.,  2012, \mn@doi
  [\apjl] {10.1088/2041-8205/748/1/L12}, \href
  {http://adsabs.harvard.edu/abs/2012ApJ...748L..12R} {748, L12}

\bibitem[\protect\citeauthoryear{{Ruderman} \& {Sutherland}}{{Ruderman} \&
  {Sutherland}}{1975}]{ruderman1975}
{Ruderman} M.~A.,  {Sutherland} P.~G.,  1975, \mn@doi [\apj] {10.1086/153393},
  \href {http://adsabs.harvard.edu/abs/1975ApJ...196...51R} {196, 51}

\bibitem[\protect\citeauthoryear{{Ruderman}, {Zhu}  \& {Chen}}{{Ruderman}
  et~al.}{1998}]{Ruderman1998}
{Ruderman} M.,  {Zhu} T.,   {Chen} K.,  1998, \mn@doi [\apj] {10.1086/305026},
  \href {http://adsabs.harvard.edu/abs/1998ApJ...492..267R} {492, 267}

\bibitem[\protect\citeauthoryear{{Schnitzeler}}{{Schnitzeler}}{2012}]{M2}
{Schnitzeler} D.~H.~F.~M.,  2012, \mn@doi [\mnras]
  {10.1111/j.1365-2966.2012.21869.x}, \href
  {http://adsabs.harvard.edu/abs/2012MNRAS.427..664S} {427, 664}

\bibitem[\protect\citeauthoryear{{Shabaltas} \& {Lai}}{{Shabaltas} \&
  {Lai}}{2012}]{sl12}
{Shabaltas} N.,  {Lai} D.,  2012, \mn@doi [\apj] {10.1088/0004-637X/748/2/148},
  \href {http://adsabs.harvard.edu/abs/2012ApJ...748..148S} {748, 148}

\bibitem[\protect\citeauthoryear{{Szary}}{{Szary}}{2013}]{szary_thesis}
{Szary} A.,  2013, preprint, \href
  {http://adsabs.harvard.edu/abs/2013arXiv1304.4203S} {} (\mn@eprint {arXiv}
  {1304.4203})

\bibitem[\protect\citeauthoryear{{Szary}, {Zhang}, {Melikidze}, {Gil}  \&
  {Xu}}{{Szary} et~al.}{2014}]{Szary2014}
{Szary} A.,  {Zhang} B.,  {Melikidze} G.~I.,  {Gil} J.,   {Xu} R.-X.,  2014,
  \mn@doi [\apj] {10.1088/0004-637X/784/1/59}, \href
  {http://adsabs.harvard.edu/abs/2014ApJ...784...59S} {784, 59}

\bibitem[\protect\citeauthoryear{{Szary}, {Melikidze}  \& {Gil}}{{Szary}
  et~al.}{2015}]{szary2015}
{Szary} A.,  {Melikidze} G.~I.,   {Gil} J.,  2015, \mn@doi [\mnras]
  {10.1093/mnras/stu2622}, \href
  {http://adsabs.harvard.edu/abs/2015MNRAS.447.2295S} {447, 2295}

\bibitem[\protect\citeauthoryear{{Taylor} \& {Cordes}}{{Taylor} \&
  {Cordes}}{1993}]{tc93}
{Taylor} J.~H.,  {Cordes} J.~M.,  1993, \mn@doi [\apj] {10.1086/172870}, \href
  {http://adsabs.harvard.edu/abs/1993ApJ...411..674T} {411, 674}

\bibitem[\protect\citeauthoryear{{Tetzlaff}, {Neuh{\"a}user}, {Hohle}  \&
  {Maciejewski}}{{Tetzlaff} et~al.}{2010}]{2010MNRAS.402.2369T}
{Tetzlaff} N.,  {Neuh{\"a}user} R.,  {Hohle} M.~M.,   {Maciejewski} G.,  2010,
  \mn@doi [\mnras] {10.1111/j.1365-2966.2009.16093.x}, \href
  {http://adsabs.harvard.edu/abs/2010MNRAS.402.2369T} {402, 2369}

\bibitem[\protect\citeauthoryear{{Timokhin}}{{Timokhin}}{2010}]{Timokhin_2010}
{Timokhin} A.~N.,  2010, \mn@doi [\mnras] {10.1111/j.1365-2966.2010.17286.x},
  \href {http://adsabs.harvard.edu/abs/2010MNRAS.408.2092T} {408, 2092}

\bibitem[\protect\citeauthoryear{{Timokhin} \& {Arons}}{{Timokhin} \&
  {Arons}}{2013}]{Timokhin_2013}
{Timokhin} A.~N.,  {Arons} J.,  2013, \mn@doi [\mnras] {10.1093/mnras/sts298},
  \href {http://adsabs.harvard.edu/abs/2013MNRAS.429...20T} {429, 20}

\bibitem[\protect\citeauthoryear{{Timokhin} \& {Harding}}{{Timokhin} \&
  {Harding}}{2015}]{Timokhin_2015}
{Timokhin} A.~N.,  {Harding} A.~K.,  2015, \mn@doi [\apj]
  {10.1088/0004-637X/810/2/144}, \href
  {http://adsabs.harvard.edu/abs/2015ApJ...810..144T} {810, 144}

\bibitem[\protect\citeauthoryear{{Torres-Forn{\'e}}, {Cerd{\'a}-Dur{\'a}n},
  {Pons}  \& {Font}}{{Torres-Forn{\'e}} et~al.}{2016}]{TorresForne2016}
{Torres-Forn{\'e}} A.,  {Cerd{\'a}-Dur{\'a}n} P.,  {Pons} J.~A.,   {Font}
  J.~A.,  2016, \mn@doi [\mnras] {10.1093/mnras/stv2926}, \href
  {http://adsabs.harvard.edu/abs/2016MNRAS.456.3813T} {456, 3813}

\bibitem[\protect\citeauthoryear{{Turolla}}{{Turolla}}{2009}]{turolla2009}
{Turolla} R.,  2009, in {Becker} W.,  ed.,  Astrophys. and Space Sci. Libr.
  Vol. 357, Astrophys. and Space Sci. Libr.. p.~141,
  \mn@doi{10.1007/978-3-540-76965-1_7}

\bibitem[\protect\citeauthoryear{{Urpin}, {Chanmugam}  \& {Sang}}{{Urpin}
  et~al.}{1994}]{urpin1994}
{Urpin} V.~A.,  {Chanmugam} G.,   {Sang} Y.,  1994, \mn@doi [\apj]
  {10.1086/174687}, \href {http://adsabs.harvard.edu/abs/1994ApJ...433..780U}
  {433, 780}

\bibitem[\protect\citeauthoryear{{Vigan{\`o}}}{{Vigan{\`o}}}{2013}]{vigano_thesis}
{Vigan{\`o}} D.,  2013, PhD thesis, University of Alicante

\bibitem[\protect\citeauthoryear{{Vigan{\`o}} \& {Pons}}{{Vigan{\`o}} \&
  {Pons}}{2012}]{vigano2012}
{Vigan{\`o}} D.,  {Pons} J.~A.,  2012, \mn@doi [\mnras]
  {10.1111/j.1365-2966.2012.21679.x}, \href
  {http://adsabs.harvard.edu/abs/2012MNRAS.425.2487V} {425, 2487}

\bibitem[\protect\citeauthoryear{{Vigan{\`o}}, {Pons}  \&
  {Miralles}}{{Vigan{\`o}} et~al.}{2012}]{2012CoPhC.183.2042V}
{Vigan{\`o}} D.,  {Pons} J.~A.,   {Miralles} J.~A.,  2012, \mn@doi [Comput.
  Phys. Communications] {10.1016/j.cpc.2012.04.029}, \href
  {http://adsabs.harvard.edu/abs/2012CoPhC.183.2042V} {183, 2042}

\bibitem[\protect\citeauthoryear{{Vigan{\`o}}, {Rea}, {Pons}, {Perna},
  {Aguilera}  \& {Miralles}}{{Vigan{\`o}} et~al.}{2013}]{vigano2013}
{Vigan{\`o}} D.,  {Rea} N.,  {Pons} J.~A.,  {Perna} R.,  {Aguilera} D.~N.,
  {Miralles} J.~A.,  2013, \mn@doi [\mnras] {10.1093/mnras/stt1008}, \href
  {http://adsabs.harvard.edu/abs/2013MNRAS.434..123V} {434, 123}

\bibitem[\protect\citeauthoryear{{Wareing} \& {Hollerbach}}{{Wareing} \&
  {Hollerbach}}{2009}]{wareing2009}
{Wareing} C.~J.,  {Hollerbach} R.,  2009, \mn@doi [\aap]
  {10.1051/0004-6361/200913452}, \href
  {http://adsabs.harvard.edu/abs/2009A%26A...508L..39W} {508, L39}

\bibitem[\protect\citeauthoryear{{Yakovlev} \& {Pethick}}{{Yakovlev} \&
  {Pethick}}{2004}]{yp2004}
{Yakovlev} D.~G.,  {Pethick} C.~J.,  2004, \mn@doi [\araa]
  {10.1146/annurev.astro.42.053102.134013}, \href
  {http://adsabs.harvard.edu/abs/2004ARA%26A..42..169Y} {42, 169}

\bibitem[\protect\citeauthoryear{{Young}, {Manchester}  \& {Johnston}}{{Young}
  et~al.}{1999}]{young1999}
{Young} M.~D.,  {Manchester} R.~N.,   {Johnston} S.,  1999, \mn@doi [\nat]
  {10.1038/23650}, \href {http://adsabs.harvard.edu/abs/1999Natur.400..848Y}
  {400, 848}

\bibitem[\protect\citeauthoryear{{de Luca}}{{de Luca}}{2008}]{deluca2008}
{de Luca} A.,  2008, in {Bassa} C.,  {Wang} Z.,  {Cumming} A.,   {Kaspi} V.~M.,
   eds,  American Inst. of Phys. Conf. Series Vol. 983, 40 Years of Pulsars:
  Millisecond Pulsars, Magnetars and More. pp 311--319 (\mn@eprint {arXiv}
  {0712.2209}), \mn@doi{10.1063/1.2900173}

\makeatother
\end{thebibliography}

\appendix

\section{Initial conditions for crust-confined magnetic fields}\label{s:appendixICs}
We impose magnetic field initial conditions which follow the logic of \cite{Aguilera2008}, but here we expand in more detail for the interested reader. In azimuthally-symmetric spherical 2D ($\partial_{\phi}\rightarrow 0$), the magnetic field may be decomposed into poloidal and toroidal components:

\begin{equation}
\vec B = \vec B_\mathrm{pol}\left(r,\theta\right) + \vec B_\mathrm{tor}\left(\phi\right)
\end{equation}

\noindent Each can then be written in terms of stream functions $\mathcal{S}(r,\theta, t)$ and $\mathcal{T}(r,\theta, t)$:

\begin{equation}
\vec B_\mathrm{pol} = \vec \nabla \times (\vec r \times \vec \nabla \mathcal S)
\end{equation}
\begin{equation}
\vec B_\mathrm{tor} = - \vec r \times \vec \nabla \mathcal{T}. 
\end{equation}

\noindent We are free to decompose $\mathcal{S}$ and $\mathcal{T}$ in terms of Legendre polynomials, and search for stationary solutions such that

\begin{equation}
\mathcal S = \sum _l C_l \frac{P_l (\cos\theta)}{r} \mathcal{S}_l(r)
\end{equation}

\noindent where $C_l$ are normalization constants ($\mathcal{T}$ has an identical form). We can then write each magnetic field component as

\begin{equation}
B_r = -\frac{1}{r^2} \sum_l C_l \mathcal{S}_l l (l+1) P_l
\end{equation}
\begin{equation}
B_\theta = - \frac{1}{r} \sum_l C_l P_l' \frac{d\mathcal{S}_l}{dr}
\end{equation}
\begin{equation}
B_\phi = - \frac{1}{r} \sum_l C_l \mathcal{T}_l P_l'
\end{equation}

\noindent To determine the radial eigenmodes, we choose so-called force-free initial conditions ($\vec{J}\times\vec{B}=0$), such that the magnetic field components obey

\begin{equation}
\nabla \times \vec B = \mu_l \vec B
\label{e:force_free}
\end{equation}

\noindent at $t=0$, where $\mu_l$ is the scale length for the radial Stokes functions of order $l$.  The choice of force free condition is simply a convenient method for imposing multipolar structure, but the NS will immediately drift away from such a configuration due to e.g. the presence of density and conductivity gradients in the induction equation. From the radial component of Eq.\,(\ref{e:force_free}), one obtains

\begin{equation}
\mu_l \mathcal S_l l(l+1) P_l = \mathcal T_l P_l' \cot \theta + \mathcal T_l P_l''.
\label{e:r_force_free}
\end{equation}
\noindent and it is immediately evident that  
\begin{equation}
\mu_l \mathcal S_l (r) = -\mathcal T_l (r).
\label{e:prop}
\end{equation}


Now considering the azimuthal component of Eq.\,(\ref{e:force_free}) we obtain the Bessel-Riccati differential equation for the poloidal Stokes function





\begin{equation}
x^2 \frac{d^2\mathcal{S}_l}{dx^2} + (x^2 - l(l+1)) \mathcal{S}_l= 0
\end{equation}
where $x=\mu_l r$ is a scaled radial coordinate. The general solutions are well-known, and for a given $l$ have the form

\begin{equation}
\Gamma_l(x) = a_l x \cdot j_l(x) + b_l x \cdot n_l(x)
\end{equation}

\noindent where $a_l$, $b_l$ are normalization coefficients, and the functions $j_l(x)$ and $n_l(x)$ can be written using Rayleigh's formula:

\begin{equation}
j_l (x) = (-x)^l \left( \frac{1}{x} \frac{\partial }{\partial x} \right) ^l \left[\frac{\sin x}{x}\right]
\end{equation}

\begin{equation}
n_l (x) = - (-x)^l \left( \frac{1}{x} \frac{\partial }{\partial x} \right) ^l \left[ \frac{\cos x}{x} \right]
\end{equation}


We can write the spherical Bessel functions in a more useful form:

\begin{equation}
j_l (x) = A_l (x) \sin(x) + \beta_l (x) \cos (x)
\end{equation}
\begin{equation}
n_l (x) = -A_l (x) \cos(x) + \beta_l (x) \sin (x)
\end{equation}

\noindent where $A_l(x)$ and $\beta_l(x)$ are polynomials in $x$, which we extract from a Fortran library. 

The field must satisfy both the vacuum outer boundary condition and the inner superconducting boundary condition. The simplest choice for the surface boundary condition is that $\mathcal{S}_l(\mu_l R_\mathrm{NS}) = 1$, which is satisfied if we choose $a_l$ and $b_l$ as
\begin{equation}
a_l = \frac{\cos (\mu_l R_\mathrm{NS})}{\mu_l R_\mathrm{NS}\cdot \beta_l (\mu_l R_\mathrm{NS})}
\label{e:a_l}
\end{equation} 
\begin{equation}
b_l = \frac{\sin (\mu_l R_\mathrm{NS})}{\mu_l R_\mathrm{NS}\cdot \beta_l (\mu_l R_\mathrm{NS})}.
\label{e:b_l}
\end{equation} 

\noindent The inner boundary condition requires $\mathcal{S}_l(\mu_l R_c) = 0$, where $R_c$ is the NS core radius. Satisfying this boundary condition then requires that
\begin{equation}
\tan [\mu_l (R_c - R_{NS})] = - \frac{\beta_l (\mu R_c)}{A_l (\mu R_c)}.
\label{e:mu_l}
\end{equation}

We employ a modified Newton's method to solve Eq.\,(\ref{e:mu_l}) numerically to obtain $\mu_l$, and then compute the coefficients $a_l$ and $b_l$ from equations Eqs.\,(\ref{e:a_l}, \ref{e:b_l}). Finally, we construct the solution $\mathcal{S}_l(x)$, and the magnetic field components have the following final forms:

\begin{equation}\label{eq:finalbr}
B_r = -\frac{1}{r^2} \sum_l l(l+1) C_l \mathcal{S}_l(x) P_l (\cos\theta)
\end{equation}
\begin{equation}\label{eq:finalbth}
B_\theta = -\frac{1}{r} \sum_l  \mu_l C_l \frac{d\mathcal{S}_l(x)}{dx} P_l' (\cos\theta)
\end{equation}
\begin{equation}\label{eq:finalbphi}
B_\phi = +\frac{1}{r} \sum_l \mu_l C_l \mathcal{S}_l(x) P_l' (\cos\theta).
\end{equation}

\end{document}